\documentclass[fleqn,usenatbib]{mnras}

\usepackage{newtxtext,newtxmath}
\usepackage[T1]{fontenc}

\usepackage[utf8]{inputenc}
\usepackage{xcolor}
\usepackage{makecell}
\newcommand{\pdv}[2]{\frac{\partial #1}{\partial #2}} % Use \pdv{}{something} for empty numerator
\usepackage{siunitx} % For numbers in scientific notation
\usepackage[percent]{overpic} % To overlay (a) and (b) on subfigures without editing the image files. Might be removable in final version.
\usepackage{mleftright} % \left and \right adds extra space. \mleft and \mright from this package do not.

\usepackage{mathtools}

\usepackage[shortcuts, acronym]{glossaries-extra}
\usepackage{glossaries-prefix}
\setabbreviationstyle[acronym]{long-short}
\glsdisablehyper

\newacronym{PIC}{PIC}{particle-in-cell}
\newacronym{NTPA}{NTPA}{nonthermal particle acceleration}
\newacronym[prefixfirst={a\ }, prefix={an\ }]{FP}{FP}{Fokker-Planck}
\newacronym{MHD}{MHD}{magnetohydrodynamic}
\newacronym{AGN}{AGN}{active galactic nuclei}
\newacronym{PWN}{PWN}{pulsar wind nebulae}
\newacronym{GRB}{GRB}{gamma ray bursts}
\newacronym{3D}{3D}{three-dimensional}
\newacronym{OU}{OU}{Ornstein–Uhlenbeck}
\newcommand{\OU}{Ornstein–Uhlenbeck} % {\ac{OU}} or {Ornstein–Uhlenbeck} if it is only used maybe 2-3 times.

\definecolor{red}{HTML}{EF559F}

\renewcommand\cite\citep

\newcommand{\Nppc}{N_\textnormal{ppc}} % number of particles per cell
\newcommand{\tFinal}{t_f}

\newcommand{\pdt}{\partial_t}
\newcommand{\pdx}{\partial_\gamma}
\newcommand{\alphaDot}{\dot{\alpha}}
\newcommand{\KdotOnK}{\dot{K}/K}
\newcommand{\gStar}{\gamma^*}

\newcommand{\rms}{\textnormal{rms}}
\newcommand{\gammaAvg}{\gamma_\textnormal{avg}} % this is used for the global average energy whereas \binMean is the bin mean energy, maybe should separate \avg command so that it can be used in over dot for the time derivative
\newcommand{\gammaPeak}{\gamma_\textnormal{pk}} % was \gamma_\textnormal{peak}
\newcommand{\gammaMax}{\gamma_\textnormal{max}}
\newcommand{\initAvgEnergy}{{\gamma_\textnormal{avg}}_0} % was {\gamma_\textnormal{avg0}} was {\overline{\gamma}_\init}
\newcommand{\pkMult}{\bcGamma / \gammaPeak}

\newcommand{\alphaDef}{-\partial \log{f} / \partial \log{\gamma}}
\newcommand{\alphaLocal}{\alpha_\textnormal{loc}}
\newcommand{\alphaSV}{\alpha}
\newcommand{\alphaFinal}{\alpha_\textnormal{fin}}

\newcommand{\alphaZero}{\alpha_\infty}
\newcommand{\DeltaAlpha}{\Delta_\alpha}
\newcommand{\tauAlpha}{\tau}

\newcommand{\Dpp}{D_{pp}}
\newcommand{\Ap}{A_p}

\newcommand{\binMean}{\overline{\gamma}}
\newcommand{\stdev}{\delta\gamma_\rms}

\newcommand{\varEnergy}{\overline{\delta\gamma^{\smash{2}}}}
\makeatletter
\AtBeginDocument{
    \check@mathfonts
    \newdimen\savedimen
    \savedimen=\fontdimen15\textfont2
    \fontdimen15\textfont2=0.7\fontdimen15\textfont2
}
\makeatother
\AtEndDocument{
    \fontdimen15\textfont2=\savedimen
}

\newcommand{\bcGamma}{\gamma_0}
\newcommand{\changeT}{\Delta t}

\newcommand{\relStdev}{\stdev / \bcGamma}
\newcommand{\relMeanChange}{\Delta\binMean / \bcGamma}

\newcommand{\me}{m} % was {m_e}
\newcommand{\restEnergy}{\me c^2}

\newcommand{\MaxwellJuttner}{Maxwell-J\"{u}ttner}
\newcommand{\Alfven}[1]{Alfv\'en{#1}}

\newcommand{\vA}{v_{\mkern-1mu \smash{A}}}
\newcommand{\uA}{u_{\smash{A}}}
\newcommand{\vao}{{\vA}_0}
\newcommand{\tva}{\vao t/L}
\newcommand{\cdtl}{c \changeT / L}

\newcommand{\Brms}{B_\rms}
\newcommand{\reo}{{\rho_e}_0} % was {\rho_{e0}}
\newcommand{\deo}{{d_e}_0}
\newcommand{\relLar}{L/\reo}

\newcommand{\deEqn}{({\gammaAvg \restEnergy / 4 \pi n_0 e^2})^{1/2}}
\newcommand{\reEqn}{\gammaAvg \restEnergy / e \Brms}

\renewcommand{\vec}[1]{\mathbf{#1}}
\newcommand\unitvec[1]{\vec{\hat{#1}}}

\newcommand{\ExB}{\vec{E} \times \vec{B}}
\newcommand{\Dpropxx}{D \propto\nobreak \gamma^2}

\newcommand{\sigmaScanVals}{3/2, 3/4,..., 3/128} % seven values with ellipsis
\newcommand{\sigmaFourVals}{3/2,\allowbreak 3/8,\allowbreak 3/32,\allowbreak 3/128}
\newcommand{\sigmaThreeVals}{3/2, 3/16, 3/128}
\newcommand{\relSizeVals}{1024,\allowbreak 683,\allowbreak 512,\allowbreak 341,\allowbreak 256,\allowbreak 171}

\newcommand{\binInitTime}{\vao t_0/L =\nobreak 6} % {\vao t_0/L = 6} or {$t_0 = 6L/\vao$}

\newcommand{\logten}{\log_{10}}

\newcommand{\dn}{dn}

\newcommand{\kd}{k_d} % {k_f}
\newcommand{\Ld}{L_d} % {L_f}

\newcommand{\speq}{\mathord{=}} % When we use e.g. N = 768, it is really being used as a name or a word rather than an equation, and less space around the equals helps it be read that way. \mathbin{=} is a bit more space but not as much as normal equation. Or if unacceptable can always change back to plain =. speq stands for sim params equals.
\newcommand{\medSigScan}{$N \speq 768$ $\sigma_0$-scan}
\newcommand{\smlSigScan}{$N \speq 384$ $\sigma_0$-scan}
\newcommand{\figXtTimes}{$t_0 \approx \{3, 6, 9\} L/\vao$}

\newcommand{\OUslope}{\theta}

\title[Energy Diffusion \& Advection in Turbulence]{Energy Diffusion and Advection Coefficients in Kinetic Simulations of Relativistic Plasma Turbulence}

\author[K. W. Wong et al.]
{
Kai W. Wong$^{1}$,
Vladimir Zhdankin$^{2,3,4}$,
Dmitri A. Uzdensky$^{1,5}$,
\newauthor
Gregory R. Werner$^{1}$,
Mitchell C. Begelman$^{6,7}$
\\
$^{1}$Center for Integrated Plasma Studies, Physics Department, 390 UCB, University of Colorado, Boulder, CO 80309, USA\\
$^{2}$Department of Physics, University of Wisconsin-Madison, Madison, WI 53706\\
$^{3}$Center for Computational Astrophysics, Flatiron Institute, 162 Fifth Avenue, New York, NY 10010, USA\\
$^{4}$Department of Astrophysical Sciences, Princeton University, Peyton Hall, Princeton, NJ 08544, USA\\
$^{5}$Rudolf Peierls Centre for Theoretical Physics, Clarendon Laboratory, University of Oxford, Parks Road, Oxford OX1 3PU, UK \\
$^{6}$JILA, University of Colorado and National Institute of Standards and Technology, 440 UCB, Boulder, CO 80309, USA\\
$^{7}$Department of Astrophysical and Planetary Sciences, 391 UCB, Boulder, CO 80309, USA
}

\date{Accepted XXX. Received YYY; in original form ZZZ}

\pubyear{2025}

\begin{document}
\label{firstpage}
\pagerange{\pageref{firstpage}--\pageref{lastpage}}
\maketitle

\begin{abstract}
Turbulent, relativistic nonthermal plasmas are ubiquitous in high-energy astrophysical systems, as inferred from broadband nonthermal emission spectra.  The underlying turbulent nonthermal particle acceleration (NTPA) processes have traditionally been modelled with a Fokker-Planck (FP) diffusion-advection equation for the particle energy distribution. We test FP-type NTPA theories by performing and analysing particle-in-cell (PIC) simulations of turbulence in collisionless relativistic pair plasma.  By tracking large numbers of particles in simulations with different initial magnetisation and system size,  we first test and confirm the applicability of the FP framework.  We then measure the FP energy diffusion ($D$) and advection ($A$) coefficients as functions of particle energy $\gamma m c^2$,  and compare their dependence to theoretical predictions.  At high energies,  we robustly find $D \sim \gamma^2$ for all cases.  Hence,  we fit $D = D_0 \gamma^2$ and find a scaling consistent with $D_0 \sim \sigma^{3/2}$ at low instantaneous magnetisation $\sigma(t)$,  flattening to $D_0 \sim \sigma$ at higher $\sigma \sim 1$.  We also find that the power-law index $\alpha(t)$ of the particle energy distribution converges exponentially in time.  We build and test an analytic model connecting the FP coefficients and~$\alpha(t)$,  predicting $A(\gamma) \sim \gamma \log \gamma$.  We confirm this functional form in our measurements of $A(\gamma,t)$,  which allows us to predict $\alpha(t)$ through the model relations.  Our results suggest that the basic second-order Fermi acceleration model,  which predicts $D_0 \sim \sigma$,  may not be a complete description of NTPA in turbulent plasmas.  These findings encourage further application of tracked particles and FP coefficients as a diagnostic in kinetic simulations of various astrophysically relevant plasma processes like collisionless shocks and magnetic reconnection.
\end{abstract}


\glsresetall
\section{Introduction}
% Astrophysical list
Relativistic charged particles with nonthermal power-law energy distributions are inferred, from observed nonthermal radiation spectra, to exist in diverse astrophysical systems such as pulsar wind nebulae (PWN), black-hole accretion flows and jets, e.g., in active galactic nuclei (AGN), and gamma-ray bursts~(GRBs).
The question of how such \ac{NTPA} occurs is a long-standing fundamental physics problem, and has been approached both through analytical theory and numerical simulations.
Many of these astrophysical systems are manifestly turbulent, and dissipation of magnetised plasma turbulence is an attractive mechanism for explaining nonthermal power-law tails of the particle energy distribution. If NTPA predominantly comes from the ideal motional electric field, 
then particle acceleration requires vigorous plasma motion. Turbulence provides this motion at a range of length-scales through the turbulent cascade, which naturally parallels a nonthermal spectrum of particle energies with corresponding resonant gyro-scales.

% Theory
While proposed NTPA mechanisms include shocks, magnetic reconnection, and turbulence, in all of these cases the predominant theoretical picture is one in which a particle's interactions with scattering structures such as shock fronts and turbulent eddies cause its energy to accumulate in a biased random walk.
The resulting energisation is captured through the \ac{FP} diffusion-advection equation in momentum space:
\begin{equation}
\pdv{F}{t}
= \pdv{}{\boldsymbol{p}} \cdot \left({\bf D}_{pp} \cdot \pdv{F}{\boldsymbol{p}}\right)
- \pdv{}{\boldsymbol{p}} \cdot (\boldsymbol{A}_p F)
,
\label{eqn:fpVec}
\end{equation}
where $F(\boldsymbol{p},t)$ is the (spatially averaged) particle momentum distribution, and ${\bf D}_{pp}(\boldsymbol{p},t)$ and $\boldsymbol{A}_p(\boldsymbol{p},t)$ are, respectively, the momentum-space diffusion coefficient tensor and advection coefficient vector.
Existing theoretical \ac{NTPA} models are usually based on quasilinear theory (QLT) \citep[e.g.,][]{kulsrud_ferrari_1971, schlickeiser1989CosmicrayTransportAcceleration, chandran_2000, demidem2020ParticleAccelerationRelativistic}, and differences between \ac{NTPA} theories are encapsulated in their predictions for the \ac{FP} coefficients.
These \ac{FP} models of particle acceleration are used widely in space physics and astrophysics
\citep[e.g.,][]{miller1990StochasticFermiAcceleration, nayakshin1998SelfconsistentFokkerPlanckTreatment, blasi2000StochasticAccelerationNonthermal, summers2000ModelGeneratingRelativistic, becker2006TimedependentStochasticParticle, liu2009CombinedModelingAcceleration, mertsch2011FermiGammaRayBubbles, asano2014TimedependentModelsBlazar, kimura2015NeutrinoCosmicRayEmission, bian2014FormationKappadistributionAccelerated, kundu_etal_2021}.
Numerical simulations offer the opportunity to first test the diffusive nature of particle energisation, and to then measure the \ac{FP} coefficients to constitute a test of the underlying theory. However, such computational analysis has not yet been conducted extensively, with just a handful of preliminary pioneering studies published recently (as described below). 

% Simulations
Magnetised plasma turbulence has been studied both with \ac{MHD} simulations and, more recently, with kinetic simulations.
To study particle acceleration in \ac{MHD} simulations, test particles are inserted and their trajectories are followed through the simulated fields \citep[e.g.,][]{dmitruk_etal_2003, dmitruk_etal_2004, kowal_etal_2012, lynn_etal_2013, lynn_etal_2014, kimura_etal_2016,kimura_etal_2019, medina-torrejon_etal_2021,sun_bai_2021, zhang_xiang_2021, bresci_etal_2022}. This method is also applied to cases where the fields are prescribed and not simulated \citep[e.g.,][]{demidem2020ParticleAccelerationRelativistic, vlahos_etal_2004, isliker2017ParticleAccelerationFractional}.
While less computationally expensive than fully kinetic simulations, this approach suffers from serious drawbacks such as arbitrary particle injection and non-self-consistent particle motion.
Kinetic effects could change the acceleration process qualitatively, by influencing the injection of particles into the acceleration process, and back-reaction of energetic particles on the electromagnetic fields. Indeed, in extreme cases, the turbulent cascade may be damped by nonthermal particles across a range of scales \citep{lemoine_etal_2024}.
In contrast to MHD, first-principles \ac{PIC} kinetic simulations naturally provide a unique capability for detailed diagnostics through the complete self-consistent history of a particle's trajectory in phase space.
However, such simulations are still relatively recent, and initial analysis has been focused on demonstrating the presence of \ac{NTPA} through largely global diagnostics such as the existence of a nonthermal power-law particle energy distribution \citep[e.g.,][]{zhdankin2017KineticTurbulenceRelativistic, zhdankin2018SystemsizeConvergenceNonthermal, comisso_sironi_2018, comisso2019InterplayMagneticallyDominated,  hankla_etal_2022, vega_etal_2022, nattila_beloborodov_2022, meringolo_etal_2023, vega_etal_2024}. 
Hence there is an opportunity for more detailed tests of \ac{NTPA} theories through direct inspection of tracked particles.

% Our previous work
Our previous paper, \citet{wong2020FirstprinciplesDemonstrationDiffusiveadvective}, a numerical analysis of an ensemble of tracked particles in a large \ac{3D} \ac{PIC} simulation,
established for the first time that the \ac{FP} framework was indeed suitable for modelling \ac{NTPA} in a kinetic simulation of driven relativistic pair-plasma turbulence.
We found that the simple diffusion-advection model works even in fully kinetic simulations of strong turbulence, providing firm first-principles computational evidence of the validity of \ac{FP}-type particle acceleration models. Subsequent work by \citet{comisso2019InterplayMagneticallyDominated} provided further evidence for diffusive acceleration in PIC simulations (of decaying relativistic turbulence) and found that the diffusion coefficient scales with magnetization in a way that is consistent with a second-order Fermi acceleration process.

% This paper
In this paper, we further test the \ac{FP} model and conduct a detailed study of time-dependent \ac{FP} energy diffusion and advection coefficients in PIC simulations of turbulent relativistic pair plasma, while varying the physical system parameters of the initial plasma magnetisation and the simulation box size relative to the initial average Larmor radius.
Since the plasma magnetization and the average Larmor radius (and hence the scale separation between the macroscopic driving scale and the plasma kinetic microscales) evolve in time in our nonradiative simulations due to continuous turbulent heating, we can investigate the dependence of the \ac{FP} coefficients not only on the initial values of these physical parameters, but also on their time-dependent instantaneous values.
In addition, we measure the power-law index of the nonthermal part of the particle energy distribution as a function of time, and relate it to the \ac{FP} coefficients with an analytical model. 
We thereby uncover how the power-law index depends on both initial and instantaneous parameters, which yields a more complete understanding of how nonthermal distributions come about in these simulations.
These insights, in turn, inform \ac{FP}-based models of NTPA in space and astrophysical systems, including those used in conjunction with global MHD simulations, and thus have important observational implications. In particular, they provide a solid, first-principles physics basis for formulating concrete usable prescriptions for spectral modelling of relativistic plasma environments around neutron stars and black holes, such as pulsar wind nebulae and black-hole jets and coronae.

% Table of contents
The paper is organised as follows.
\autoref{sec:theory} outlines previous analytical research on turbulent \ac{NTPA}, and presents an analytic model relating the \ac{FP} coefficients to the evolution of the power-law index of the particle energy distribution.
\autoref{sec:sims} describes our \ac{PIC} simulation setup and parameter scans.
\autoref{sec:global} discusses general features of the turbulent particle energisation and their time evolution in our simulations.
\autoref{sec:alpha} presents our measurements of the power-law index of the particle energy distribution as a function of time and the scalings of the key parameters describing this evolution with the dimensionless system parameters.
\autoref{sec:fp} tests whether particle energies can be modelled as diffusive, and whether the measured \ac{FP} coefficients reproduce the evolution of the particle energy distribution from the \ac{PIC} simulations.
\autoref{sec:dCoeff} presents our findings on the energy diffusion coefficient as a function of time and system parameters.
\autoref{sec:aCoeff} shows measurements of the energy advection coefficient as a function of time and system parameters, and examines how it relates to the theoretical model described in \autoref{sec:theory}.
Finally, \autoref{sec:conc} summarises our results.

%\newpage
\section{Theory}
\label{sec:theory}
\subsection{Fokker-Planck theories of nonthermal particle acceleration} \label{sec:FPeq}
The full momentum-space \ac{FP} equation is given in \eqref{eqn:fpVec}. There, the diffusion and advection coefficients are, respectively, a tensor (${\bf D}_{pp}$) and a vector ($\boldsymbol{A}_p$), and encapsulate all possibilities of diffusion and advection of the particle  distribution between different components of momentum (including, e.g., pitch-angle scattering). In this paper, we will now assume the form $\boldsymbol{A}_p=A_p \hat{\boldsymbol{p}}$ and ${\bf D}_{pp}=D_{pp} {\bf I}$, where $A_p$ and $D_{pp}$ are the scalar advection and diffusion coefficients, respectively, with $\hat{\boldsymbol{p}}$ being the unit vector along $\boldsymbol{p}$ and ${\bf I}$ the identity matrix.

Furthermore, we will limit the analysis to a reduced description of the \ac{FP} equation for the global particle distribution $f(\gamma, t)$ in energy only, neglecting the pitch-angle dependence:
\begin{equation}
\pdt f = \pdx(D \pdx f - A f),
\label{eqn:fpEne}
\end{equation}
where the particle density is $\dn = f(\gamma) d\gamma$, and the energy diffusion and advection coefficients $D$ and $A$ are scalar functions of particle energy~$\gamma m c^2$, where $\me$ is the particle rest-mass and $c$ is the speed of light.
However, since it is common to state the \ac{FP} coefficients in momentum space, it is useful to provide, for reference, conversion formulae between the energy-space and momentum-space forms.
Beginning with \eqref{eqn:fpVec}, where the particle density $\dn = F(\vec{p}) d^3\vec{p}$, and assuming for simplicity that $F$, $\Dpp$, and~$\Ap$ are isotropic (i.e., depend only on $p \equiv |\vec{p}|$, thus neglecting any pitch-angle dependence), we obtain
\begin{equation}
\pdv{F}{t} = \frac{1}{p^2} \pdv{}{p} \left(p^2 \Dpp \pdv{F}{p} \right) - \frac{1}{p^2} \pdv{}{p} (p^2 \Ap F)
,
\end{equation}
where $\dn$ simplifies to~$4\pi p^2 F(p)\, dp$.
This is the form used in, e.g., \citet{schlickeiser1985ViableMechanismEstablish, schlickeiser1989CosmicrayTransportAcceleration}.
Finally, in the ultrarelativistic limit $\gamma \gg 1$ that we will focus on in this paper,  
the particle energy $\gamma mc^2$ is just directly proportional to its momentum~$p$, i.e., $\gamma = p/mc$; we can then substitute $f(\gamma) = 4\pi p^2 F(p)$ and thus write 
\begin{equation}
\pdv{f}{t} =
\pdv{}{\gamma} \left[\Dpp\left(\pdv{f}{\gamma} - 2\,\frac{f}{\gamma}\right)\right]
- \pdv{}{\gamma}(\Ap f)
.
\end{equation}
Comparing this to \eqref{eqn:fpEne}, we see that the momentum-space coefficients $\Dpp$ and $A_p$ are related to the energy-space coefficients (without subscript) by $D = \Dpp$ and $A = \Ap + 2\Dpp / \gamma$.

\subsection{Analytical model of the nonthermal power-law tail} \label{sec:analyticalModel}
\newcommand{\fPowerLaw}{f(\gamma, t) \nobreak=\nobreak K(t) \gamma^{-\alpha(t)}}
\newcommand{\dGammaSqEqn}{D(\gamma, t) \nobreak=\nobreak D_0(t) \gamma^2}
\newcommand{\anaAdvEqn}{A =\nobreak A_0 \gamma \log(\gamma/\gStar_A)}

Previous PIC numerical simulation studies have shown the formation and gradual evolution of a relativistic nonthermal power-law section in the particle energy distribution in driven collisionless plasma turbulence \cite[e.g.,][]{zhdankin2017KineticTurbulenceRelativistic, zhdankin2018SystemsizeConvergenceNonthermal,wong2020FirstprinciplesDemonstrationDiffusiveadvective}.
That is, there is a significant time interval where a segment of $f(\gamma)$ is well approximated by a power law.
We wish to understand the relationship between the \ac{FP} coefficients and the time evolution of the power-law index of the nonthermal section of~$f(\gamma)$. Hence, we consider the \ac{FP} equation \eqref{eqn:fpEne}, where $f$, $D$, and $A$ are functions of $\gamma$ and~$t$.
We will focus here only on the nonthermal region, and only for the times when this power law is well-formed and distinct, so that the high-energy part of $f(\gamma)$ can be said to be an evolving power law 
\begin{equation}
\fPowerLaw.
\label{eqn:fPowerLaw}
\end{equation}
Here, $K$ should not be viewed as a normalisation factor, since the number of particles in the power law does not have to be conserved.
Instead, we imagine reservoirs of particles at low and high energies. At low energies, this is naturally interpreted as the thermal bulk. At high energies, this could be a form of escape; in simulation terms, this may be a high-energy pileup as described by \citet{zhdankin2017KineticTurbulenceRelativistic, zhdankin2018SystemsizeConvergenceNonthermal}. Particles moving in and out of these reservoirs can change the total number of particles in the nonthermal tail.

Theoretical models based on QLT \citep[e.g.,][]{kulsrud_ferrari_1971, melrose_1974, skilling_1975, blandford_eichler_1987, chandran_2000, cho_lazarian_2006, demidem2020ParticleAccelerationRelativistic}
and generalized Fermi acceleration \citep{lemoine_2019, lemoine_2021, lemoine_2025} commonly predict $D(\gamma, t) \propto \gamma^2$ in the nonthermal range.
Hence, for this model, we assume
\begin{equation}
\dGammaSqEqn.
\label{eqn:dGammaSq}
\end{equation}
In the absence of an advection coefficient, this diffusion coefficient causes the mean energy $\binMean$ of a collection of particles to increase at the instantaneous rate of $\dot{\binMean} \equiv \partial_t \binMean = 2D_0\binMean$: an exponential pace, in accordance with the Fermi acceleration theory.
Hence, $D_0$ may be interpreted as the inverse diffusive acceleration time, and the scaling $\Dpropxx$ implies that this characteristic acceleration time is independent of the particle energy $(\dot{\binMean}/\binMean = \textrm{const})$.

While we could also assume a functional form for~$A(\gamma,t)$, specifying three independent forms for $f$, $D$ and $A$ will almost certainly be inconsistent due to over-constraining the problem.
Instead, since $A$ is by far the most uncertain theoretically, we use \eqref{eqn:fpEne} with the above adopted functional forms \eqref{eqn:fPowerLaw} and \eqref{eqn:dGammaSq}
for $f(\gamma, t)$ and~$D(\gamma,t)$, respectively, to solve for $A(\gamma, t)$ analytically, and then later compare it to our numerical measurements of $A$ in \autoref{sec:aCoeff}. That is, we infer $A$ from consistency conditions on the existence of a power-law range in the energy distribution.

First, we integrate the \ac{FP} equation \eqref{eqn:fpEne}
from $\gamma$ to~$\infty$. 
Keeping in mind \eqref{eqn:fPowerLaw} and \eqref{eqn:dGammaSq}, we assume $\alpha > 1$ and $A f \rightarrow 0$ as $\gamma \rightarrow \infty$, so that all the upper integral limits can be discarded.
The non-advection terms then evaluate to:
\begin{equation}
\begin{split}
& \int_{\gamma}^{\infty} \partial_{\gamma'}(D \partial_{\gamma'} f) d\gamma' 
= -D \pdx f 
= -D_0 \gamma^2 (-\alpha / \gamma) f = D_0 \alpha \gamma f, \\
&\int_{\gamma}^{\infty} \! \pdt f d\gamma'
= - \pdt \left( K \frac{\gamma^{1 - \alpha}}{1 - \alpha} \right)
= - \frac{\gamma f}{1 - \alpha} \left( -\alphaDot \log\gamma + \frac{\dot{K}}{K} + \frac{\alphaDot}{1 - \alpha} \right).
\end{split}
\end{equation}
%% line by line working
% \begin{align}
% \textstyle{\int} \pdt f d\gamma &= \pdt \left( K \frac{\gamma^{1 - \alpha}}{1 - \alpha} \right) \\
% &= \frac{K}{1 - \alpha} \bigl( -\alphaDot \gamma^{1 - \alpha} \log\gamma \bigr)
%     + \gamma^{1 - \alpha} \frac{(1 - \alpha)\dot{K} + K \alphaDot}{(1 - \alpha)^2} \\
% &= \frac{\gamma f}{1 - \alpha} \left( -\alphaDot \log\gamma + \frac{\dot{K}}{K} + \frac{\alphaDot}{1 - \alpha} \right)
% \end{align}
We collect into \eqref{eqn:fpEne} and eliminate the common factors of $f$ to get
\begin{align}
% A &= - D_0 \alpha \gamma - \frac{\gamma}{1 - \alpha} \left( -\alphaDot \log\gamma + \frac{\dot{K}}{K} + \frac{\alphaDot}{1 - \alpha} \right) \\
% A &= - \frac{\gamma}{1 - \alpha} \left( -\alphaDot \log\gamma + \frac{\dot{K}}{K} + \frac{\alphaDot}{1 - \alpha} + D_0 \alpha (1 - \alpha) \right)
A &= -\, \frac{\gamma}{1 - \alpha} \left( -\alphaDot \log\gamma + \frac{\dot{K}}{K} + \frac{\alphaDot}{1 - \alpha} + D_0 (\alpha - \alpha^2) \right).
\label{eqn:Aterms}
\end{align}
This can be recast in a convenient, compact form as 
\begin{equation}
    A(\gamma,t) = A_0(t)\, \gamma\, \log \mleft[ \gamma/\gStar_A(t) \mright],
    \label{eqn:Axlogx}
\end{equation}
with:
\begin{align}
    &A_0(t)
    = \frac{\alphaDot}{1 - \alpha}
    = -\frac{d\log(\alpha - 1)}{dt}
    \label{eqn:A0}
    ,\\
    &\gStar_A(t) = \exp\mleft[ \frac{\KdotOnK + D_0 (\alpha - \alpha^2)}{\alphaDot} + \frac{1}{1 - \alpha} \mright]
    ,
    \label{eqn:xA}
\end{align}
where all right-hand-side variables in \eqref{eqn:A0} and \eqref{eqn:xA} are functions of $t$ only.
The result \eqref{eqn:Axlogx} is the prediction for the functional form of~$A(\gamma,t)$ required to maintain a power-law distribution.
The $\gamma$-dependence is remarkably simple and contained only in \eqref{eqn:Axlogx}, with \eqref{eqn:A0} and \eqref{eqn:xA} functions of $t$ alone.
As a consequence of the assumed quadratic diffusion coefficient $\Dpropxx$,
only one term in the large parentheses of \eqref{eqn:Aterms} has $\gamma$-dependence,
and this directly leads to a generic $\gamma \log \gamma$ energy dependence of the advection coefficient~$A$, as long as $\dot{\alpha} \neq 0$. If instead $\alphaDot=0$, then $A$ is simply proportional to~$\gamma$ by \eqref{eqn:Aterms}.
Since $\alphaDot$ is usually negative in the PIC turbulence simulations, with the power law tending to harden over time, $A_0$ is positive. 
Hence,~\eqref{eqn:Axlogx} predicts that $A$ is negative at low energies and positive at high energies, with the sign change occurring at~$\gStar_A$.
Furthermore, $A$ has a minimum at $\gamma=\gStar_A / e$, with value $A_\mathrm{min} = -A_0 \gStar_A/e$.
We note that at high energies, $A$ is faster than first-order Fermi acceleration (where $A \propto \gamma$) by a factor of $\log\gamma$. This is a surprising property that appears to be generic for maintaining a time-evolving power-law particle energy distribution given $D \propto \gamma^2$.
Finally, an equivalent statement to \eqref{eqn:Axlogx} is that $A/\gamma$ is linear in $\log\gamma$, which we will use in \autoref{sec:aCoeff}.

Up to this point, we have introduced fixed $\gamma$-dependencies for the particle energy distribution $f$ and \ac{FP} coefficients $D$ and $A$, parameterised by five functions of time
$K$, $\alpha$, $D_0$, $A_0$, and $\gStar_A$, which are associated to the former three quantities as follows:
\begin{description}
\item $f: K, \alpha; \quad D: D_0; \quad A: A_0, \gStar_A.$
\end{description}
While all of these ultimately source from the \ac{PIC} simulation, the particle-energy-distribution group of variables and the \ac{FP}-coefficient group are measured in two substantially different ways: the former use the overall particle energy distribution and the latter, tracked particles (see \autoref{subsec:trackOsc} for some details on tracked particles). 

We later find in \autoref{sec:aCoeff} that $A_0$ and $\gStar_A$ both fluctuate substantially in time (or at least, their measurements do), whereas their effect on the power law is cumulative over time, innately smoothing out the fluctuations; hence an integral comparison can be less noisy and more illuminating.
The tradeoff for this smoothing is that we will have to select the integration constants (somewhat arbitrarily).
To enable this, we rearrange the $A$-variables equations \eqref{eqn:A0} and \eqref{eqn:xA} into time integrals for the power-law variables $K(t)$ and $\alpha(t)$.
From~\eqref{eqn:A0},
\begin{equation}
% \alpha(t) = 1 + C e^{-\int\limits_{t_0}^{t} A_0(t') \,dt'}\, ,
% \alpha(t) = 1 + C \exp \mleft[ -\!\int_{t_0}^{t} \!A_0(t') dt' \mright],
% \alpha(t) = 1 + [\alpha(t_0) - 1] \exp \mleft[ -\!\int_{t_0}^{t} \!A_0(t') dt' \mright],
\frac{\alpha(t) - 1}{\alpha(t_0) - 1} = \exp \mleft[ -\!\int_{t_0}^{t} \!A_0(t') dt' \mright],
% \frac{\alpha(t) - 1}{\alpha(t_0) - 1} = \exp \!\int_{t_0}^{t} \!-A_0(t') dt',
\label{eqn:cInt}
\end{equation}
where $t_0$ is an arbitrary reference time and
$\alpha(t_0)$ is an integration constant. 
Similarly, \eqref{eqn:xA} rearranged for $\KdotOnK$ integrates to
\begin{equation}
\frac{K(t)}{K(t_0)} = \exp
\mleft\{
    \int_{t_0}^t \mleft[
        D_0 (\alpha^2 - \alpha) + \alphaDot \left( \log\gStar_A + \frac{1}{\alpha - 1} \right)
    \mright] \,dt'
\mright\}
.
\label{eqn:kIntA0}
\end{equation}
Here, $t_0$ is again a reference time, 
$K(t_0)$ is an integration constant and $D_0$, $\alpha$, $\alphaDot$, and $\gamma_A^*$ are all functions of time~$t'$.
One may optionally then trade $\alphaDot$ for $A_0$ using \eqref{eqn:A0}.
While theoretically equivalent, $\alphaDot$ and $A_0$ are obtained from different data ($f$ and $A$, respectively) and one or the other may be preferable for noise or other technical considerations.
We also note that the exponential form of \eqref{eqn:cInt} has a suggestive (although not quite fully compatible) similarity to the exponential fit to $\alpha(t)$ introduced in \autoref{sec:alpha}.

To compute the advection coefficient~$A(\gamma,t)$ in our PIC simulations, we will first measure the energy-dependent average rate of particle energy gain $M \equiv A + \pdx D$ (see \autoref{sec:fpTests}).
Later, we will find it convenient to compare directly between theoretical and measured $M$ rather than $A$ because $M$ does not depend on $D$ and its associated measurement noise.
Hence we derive analogues of
% (\ref{eqn:Axlogx}, \ref{eqn:A0}, \ref{eqn:xA})
(\ref{eqn:Axlogx}--\ref{eqn:xA})
for~$M$ using $\pdx D = 2 D_0 \gamma$ which comes from \eqref{eqn:dGammaSq}:
    \begin{equation}
    M(\gamma,t) = M_0 \gamma \log(\gamma/\gStar_M),
    \label{eqn:Mxlogx}
    \end{equation}
where $M_0 = A_0$ is the same as in \eqref{eqn:A0} (it is convenient to have a different symbol for this prefactor as later fits to $A$ and $M$ will not necessarily result in the same value) and
    \begin{equation}
    \gStar_M(t) = \exp\mleft[ \frac{\dot{K}/K + D_0 (-\alpha^2 + 3\alpha - 2)}{\alphaDot} + \frac{1}{1 - \alpha} \mright]
    ,
    \label{eqn:xM}
    \end{equation}
differing from $\gStar_A$ only by the factor multiplying~$D_0$.
The relationship between $\gStar_A$ and $\gStar_M$ is
$\log\gStar_A - \log\gStar_M = 2D_0 / A_0$.
The corresponding integral forms are:
\begin{align}
&\frac{\alpha(t) - 1}{\alpha(t_0) - 1} = \exp \mleft[ -\!\int_{t_0}^{t} \!M_0(t') dt' \mright]
\label{eqn:alphaIntM0}
,\\
&\frac{K(t)}{K(t_0)} = \exp
\mleft\{
    \int_{t_0}^t \mleft[
        D_0 (\alpha^2 - 3\alpha + 2) + \alphaDot \left( \log\gStar_M + \frac{1}{\alpha - 1} \right)
    \mright] \,dt'
\mright\}
\label{eqn:kIntM0}
.
\end{align}

In summary, by adopting an Ansatz that the particle energy distribution is a time-dependent power law
$\fPowerLaw$ \eqref{eqn:fPowerLaw} and that the energy diffusion coefficient is $\dGammaSqEqn$ \eqref{eqn:dGammaSq},
we obtain a prediction for the energy advection coefficient of
$A(\gamma,t) =\nobreak A_0(t)\, \gamma\, \log[\gamma/\gStar_A(t)]$ \eqref{eqn:Axlogx},
where $A_0$ is a function of $\alpha$ and~$\alphaDot$ \eqref{eqn:A0} and $\gStar_A$ is a function of $\KdotOnK$, $D_0$, $\alpha$, and~$\alphaDot$ \eqref{eqn:xA}.

Our analytical model does not explain why a power-law distribution arises, but this might be expected on general grounds from the underlying scale invariance of turbulence at MHD scales. 
For example, it may be a natural consequence of particle segregation based on acceleration rate \citep{lemoine2020PowerlawSpectraStochastic}.
The assumption of a power-law distribution is also supported by recent maximum-entropy models of particle acceleration that derive such distributions, with the index being connected to the characteristic momentum scale of irreversible dissipation \citep[e.g.,][and references therein]{zhdankin_2022}.

\section{Simulations} \label{sec:sims}
We analyse \ac{3D} \ac{PIC} simulations of externally driven turbulence in collisionless relativistic pair plasma performed with our code {\sc Zeltron} \citep{cerutti2013SimulationsParticleAcceleration}.
The system is a periodic cube of side length $L$ and a guide magnetic field~$B_0 \unitvec{z}$.
The plasma, with total (electrons and positrons combined) charged particle density~$n_0$,
is initialised from a uniform isotropic \MaxwellJuttner{} relativistic thermal distribution.
We fix the initial temperature to $T_0 \equiv \theta_0 \restEnergy =100\, \restEnergy$,
corresponding to an average Lorentz factor of $\initAvgEnergy \approx 3 \theta_0 = 300$.

We define the (time-dependent) characteristic Larmor radius $\rho_e(t) \equiv \reEqn$
and skin depth $d_e(t) \equiv \nobreak \smash{\deEqn}$,
where $\gammaAvg(t)$ is the instantaneous average Lorentz factor and $\Brms(t)$ is the rms magnetic field strength.
The corresponding initial values are denoted $\reo$ and~$\deo$.
The system then has three dimensionless physical parameters (with any choice of two independent), which are ratios of the length-scales $\rho_e$, $d_e$, and~$L$.
These are the (``hot'') magnetisation
$\smash{\sigma \equiv 3 \Brms^2 / 16\pi n_0 \gammaAvg \restEnergy = (3/4)(d_e/\rho_e)^2}$,
which is the ratio of the magnetic enthalpy $\Brms^2 / 4\pi$ to the relativistic plasma enthalpy $(4/3)n_0 \gammaAvg \restEnergy$; 
the system size relative to the Larmor radius~$L/\rho_e$; and the system size relative to the skin depth~$L/d_e$.
For the initial ultrarelativistic \MaxwellJuttner{} distribution, the initial magnetisation $\smash{\sigma_0 = B_0^2/16\pi n_0 T_0}$ corresponds to an initial plasma-to-magnetic pressure ratio of $\beta_0 = 8 \pi n_0 T_0 / B_0^2 = 1/2\sigma_0$.

Turbulence is continuously electromagnetically driven \citep{tenbarge2014OscillatingLangevinAntenna}
at wavenumbers $\kd = 2\pi / L$
and becomes fully developed after several light-crossing times; the turbulence-driving implementation and other technical simulation details are described in \citet{zhdankin2018NumericalInvestigationKinetic}. 
The driving strength is chosen so that the rms level of the turbulent magnetic fluctuations $\delta \Brms$ is comparable to the magnitude of the guide field~$B_0$.
The turbulence is essentially \Alfven{ic} \citep{zhdankin2018NumericalInvestigationKinetic},
with characteristic \Alfven{} velocity $\vA/c \equiv [\sigma/(\sigma + 1)]^{1/2}$.
% with initial \Alfven{} velocity $v_{A0}/c \equiv [\sigma_0/(\sigma_0 + 1)]^{1/2}$.

The simulation is performed on a uniform grid of $N^3$ cells.
The grid spacing $\Delta x \equiv L/N$ is chosen to resolve the smaller of $\reo$ and~$\deo$, as well as the initial Debye length ${\lambda_D}_0 = \sqrt{T_0/4\pi n_0 e^2}$ (for an ultra-relativistic \MaxwellJuttner{} distribution, $d_e / \lambda_D = \sqrt{3}$).
More precisely, we choose $\Delta x = \min(\reo, \sqrt{2}\deo) / 1.5$.
Except where otherwise stated, the combined number of electron and positron macroparticles per cell is $\Nppc = 64$.
The simulation durations $t_f$ range from about 10 to 100 light-crossing times~$L/c$ depending primarily on~$\sigma_0$ (lower $\sigma_0$ require longer runs, due to the longer principal dynamical timescale set by~$L/\vA$), with $t_f \sim 20 L/c$ being most common.

\subsection{Parameter scans and simulation list}
\newcommand{\tablefrac}{\dfrac}
\begin{table}
\caption{List of simulations and their parameters.}
\label{table:sims}
\setlength\tabcolsep{1.0pt} % This is just the minimum, the \extracolsep later instructs it to fill the full width automatically.
\small
\begin{tabular*}{\linewidth}{@{\extracolsep\fill}l|rrrrrrrr}
\hline
~\\[-2ex] % https://tex.stackexchange.com/questions/305286/how-to-avoid-equation-from-touching-h-lines-in-tables
    & $N$  & $\sigma_0$ & $\tablefrac{L}{\reo}$ & $\tablefrac{\reo}{\Delta x}$ & $\tablefrac{L}{\deo}$ & $\tablefrac{\deo}{\Delta x}$ & $\tablefrac{c\tFinal}{L}$ & $\Nppc$ \\
[-2ex]\\
\hline
$\sigma_0$ scan
        & 384  & 3/2   & 256  & 1.5 & 181 & 2.1 & 7.8    & 64  \\
$N=384$ & 384  & 3/8   & 256  & 1.5 & 362 & 1.1 & 22.3   & 64  \\
        & 384  & 3/32  & 128  & 3.0 & 362 & 1.1 & 64.9   & 64  \\
        & 384  & 3/128 & 64   & 6.0 & 362 & 1.1 & 122.3  & 64  \\
\hline                                               
$\sigma_0$ scan
        & 768  & 3/2   & 512  & 1.5 & 362 & 2.1 & 22.3   & 64  \\ 
$N=768$ & 768  & 3/4   & 512  & 1.5 & 512 & 1.5 & 22.3   & 64  \\
        & 768  & 3/8   & 512  & 1.5 & 724 & 1.1 & 22.3   & 64  \\
        & 768  & 3/16  & 362  & 2.1 & 724 & 1.1 & 20.9   & 64  \\
        & 768  & 3/32  & 256  & 3.0 & 724 & 1.1 & 55.2   & 64  \\
        & 768  & 3/64  & 181  & 4.2 & 724 & 1.1 & 59.7   & 64  \\
        & 768  & 3/128 & 128  & 6.0 & 724 & 1.1 & 117.8  & 64  \\
\hline                                            
$\Nppc$ scan%\hspace{1mm}
        & 384  & 3/8   & 256  & 1.5 & 362 & 1.1 & 22.3   & 8   \\
        & 384  & 3/8   & 256  & 1.5 & 362 & 1.1 & 22.3   & 16  \\
        & 384  & 3/8   & 256  & 1.5 & 362 & 1.1 & 22.3   & 32  \\
        & 384  & 3/8   & 256  & 1.5 & 362 & 1.1 & 22.3   & 128 \\
        & 384  & 3/8   & 256  & 1.5 & 362 & 1.1 & 22.3   & 256 \\
\hline                                               
system-size
        & 256  & 3/8   & 171  & 1.5 & 241 & 1.1 & 16.7   & 64  \\
scan        & 384  & 3/8   & 256  & 1.5 & 362 & 1.1 & 22.3   & 64  \\
        & 512  & 3/8   & 341  & 1.5 & 483 & 1.1 & 22.3   & 64  \\
        & 768  & 3/8   & 512  & 1.5 & 724 & 1.1 & 16.7   & 64  \\
        & 1024 & 3/8   & 683  & 1.5 & 965 & 1.1 & 16.7   & 64  \\
        & 1536 & 3/8   & 1024 & 1.5 & 1448& 1.1 & 14.2   & 64  \\
\hline
\end{tabular*}
\end{table}

One of the primary aims of this study is to extend the work of \citet{wong2020FirstprinciplesDemonstrationDiffusiveadvective} by analysing and comparing the results of numerous simulations with a range of initial system parameters; these simulations are summarised in \autoref{table:sims}. They are grouped as follows:
\begin{enumerate}
\item A system-size scan at fixed $\sigma_0 = 3/8$ (where $\reo = \sqrt{2}\deo$) with $L/\reo$ ranging from 171 to~1024, corresponding to $N$ varying from 256 to~1536. This group contains our largest simulation with $L/\reo =\nobreak 1024$, performed on an $N^3=1536^3$-cell grid.
\item Sets of smaller simulations with varying $\sigma_0$:
one with $N=768$ and $\sigma_0 \in \{\sigmaScanVals\}$,
and another with $N=384$ and $\sigma_0 \in \{\sigmaFourVals\}$.
As mentioned above, we set $\Delta x = \min(\reo, \sqrt{2}\deo) / 1.5$
to resolve the smaller of the two kinetic scales while
maximising the inertial range for given grid size (which is closely related to computational cost), leading to $L/\reo$ and $L/\deo$ varying as shown in the table.
\item A scan to investigate the convergence of the results with the number of particles per cell, $\Nppc$, for a representative case with $\sigma_0 = 3/8$ and $L/\reo=256$ ($N=384$).
\end{enumerate}

Let us discuss briefly some complications caused by having three initial dimensionless ratios: $\relLar$, $L/\deo$, and $\deo/\reo \propto \sigma_0^{1/2}$.
A parameter scan must select at least two of these ratios to change with at most one held fixed (as fixing two ratios would fix all three).
When aiming to isolate one variable, the handling of the other two parameters can still be consequential.
For the system-size scan, the choice is obvious: as $L$ is the privileged (always largest) scale, we fix~$\deo/\reo$.
However, the $\sigma_0$ scan is more complicated: as $\reo$ and $\deo$ have no particular ordering, it is unclear how best to set $L/\reo$ and~$L/\deo$.
For example, increasing $\sigma_0$ with fixed $L/\reo$ will decrease~$L/\deo$, and, supposing $\deo > \reo$, also decrease the separation between the driving scale and largest microscale (i.e., the inertial range).
Then, even though the change in $L/\deo$ is mathematically predetermined due to fixed $L/\reo$,
it is still important to consider carefully whether  some change in the results would be best attributed to the higher magnetisation or to the shorter inertial range.

A full two-dimensional parameter scan is unfortunately impractical. 
However, many of the results related to the inertial range are expected to become insensitive to $L/\reo$ and $L/\deo$ in the limit of $L/\reo \gg 1$ and $L/\deo \gg 1$.
Thus, the system-size scan results can provide some assurance that our simulations are in this asymptotic large-system regime and hence the effect of the varying scale hierarchies in the $\sigma_0$ scans is minor.
Nevertheless, one should keep in mind this limitation of our study with respect to the $\sigma_0$ scans.
Accordingly, we refrain from declaring primacy to one or the other of $L/\reo$ or $L/\deo$ and refer to the $\sigma_0$ scans by their fixed~$N$, which, although not a fully-fledged physical parameter, is (by the $\Delta x$ choice) almost equal to the initial ratio of $L$ to the smallest microscale.

\subsection{Tracked particles and energy oscillation removal}
\label{subsec:trackOsc}
Our analysis uses large numbers of tracked particles, for which the position, momentum, and local electromagnetic field vectors are recorded at fine enough time intervals to resolve details such as particle orbits. While memory and storage constraints prevent us from recording these data for all the $\mathbin{\sim}10^{11}$ particles in the simulations, the $\mathbin{\sim}10^6$ randomly selected tracked particles are sufficient to form a high-quality statistical ensemble representative of the overall particle distribution. 

There are large (order-unity) oscillations in particle energy due to $\ExB$ drift, particularly significant in relativistic turbulence with $E_{\rm rms}\sim B_0$.  These were described in \citet{wong2020FirstprinciplesDemonstrationDiffusiveadvective}, along with a removal procedure which is necessary to facilitate \ac{FP}-type analysis. We briefly recount these here.

\newcommand{\pref}{D}
\newcommand{\vPrefMag}{v_\pref}
\newcommand{\vPref}{\vec{v}_\pref}
\newcommand{\EsqBsq}{E^2 +\nobreak B^2}
\newcommand{\invSqrt}[1]{(#1)^{-1/2}}
\newcommand{\tLarDef}[2]{2 \pi {#1} \me c / e {#2}}
Consider a charged particle with four-velocity $\gamma \vec{v}$ moving in constant uniform electromagnetic fields.
We use unprimed variables for the lab frame and primed variables for the $\ExB$ drift frame moving with velocity~$\vPref$, given by $c\vPref/(c^2 + \vPrefMag^2) = \ExB/(\EsqBsq)$.
The particle's lab-frame energy, obtained by inverse Lorentz transform from simple gyration in the drift frame, is given by the equations
\begin{align}
&\gamma(t) = \gamma_\pref \gamma' \left[1 +
\beta_\pref \frac{v_\perp'}{c}
\cos{\left(\omega' t'\right)}\right]
\label{eqn:labGamma}
,\\
&t = t_0 + \gamma_\pref \left[t' + \beta_\pref {v_\perp' \over \omega' c } \sin{\left(\omega't'\right)} \right].
\end{align}
Here,
$t$ is the coordinate time,
$\beta_\pref =\nobreak \vPrefMag / c$,
$\gamma_\pref = \invSqrt{1 -\nobreak \beta_\pref^2}$,
$v_\perp'$ is the particle's primed-frame velocity perpendicular to $\vec{B}'$,
$\omega' =\nobreak eB'/\gamma' \me c$ is the cyclotron frequency
with $B' =\nobreak B / \gamma_\pref$,
and
$t_0$ is a phase.
Since we are considering relativistic particles ($v_\perp' \sim\nobreak c$) and relativistic turbulence ($E_\rms \sim\nobreak B_0$ and $\beta_\pref \sim\nobreak 1$),
the oscillation magnitude is comparable to~$\gamma$.
In actuality, the motion and fields are not constant, but the result essentially still applies, and such oscillations would heavily complicate \ac{FP} analysis.

\newcommand{\larmorAvg}[1]{\langle #1 \rangle}
\newcommand{\vPrefAvg}{\larmorAvg{\vPref}}
The oscillation removal procedure 
is informed by \eqref{eqn:labGamma}, which indicates
that in the idealized case of uniform constant fields,
the secular component of the lab-frame energy is~$\gamma_\pref \gamma'$.
We add a step to smooth $\vPref$ fluctuations as the particle traverses small-scale fields, which is not strictly necessary but slightly improves the result compared to just $\gamma_D \gamma'$.
First, we average $\vPref$ over the smoothed gyroperiod $\tLarDef{\gamma'}{B}$, where $\gamma' =\nobreak \gamma_D \gamma (1 - \vPref \cdot \vec{v} / c^2)$, and
denote this smoothed value by~$\vPrefAvg$.
We then define the smoothed particle energy to be
$\larmorAvg{\gamma_\pref} \larmorAvg{\gamma'} $,
where
$\larmorAvg{\gamma'} = \larmorAvg{\gamma_D} \gamma (1 - \vPrefAvg \cdot\nobreak \vec{v} / c^2)$
and
$\larmorAvg{\gamma_\pref} \equiv (1 - \vPrefAvg^2/c^2)^{-1/2}$.
Here, the Lorentz boost on $\gamma$ has been done twice: first to find a window length to smooth $\vPref$, and then again with the smoothed $\vPref$ to obtain the final result.

This procedure extracts the secular component of particle energy, greatly reducing oscillations and allowing us to test the \ac{FP} picture of \ac{NTPA}.
Hereafter, $\gamma$ and ``energy'' refer to $\larmorAvg{\gamma_\pref} \larmorAvg{\gamma'}$,
except for overall particle energy distributions and magnetic energy spectra (where the difference is minor anyway).

%\newpage
\section{System overview} \label{sec:global}
Before beginning the main analysis, we first provide a qualitative overview of how the driven turbulent system evolves. We briefly present the time evolution of the overall particle energy spectrum and average energy, and then examine a particle acceleration event superimposed on the turbulent fields.

\begin{figure}
\centering
\includegraphics[width=\linewidth]{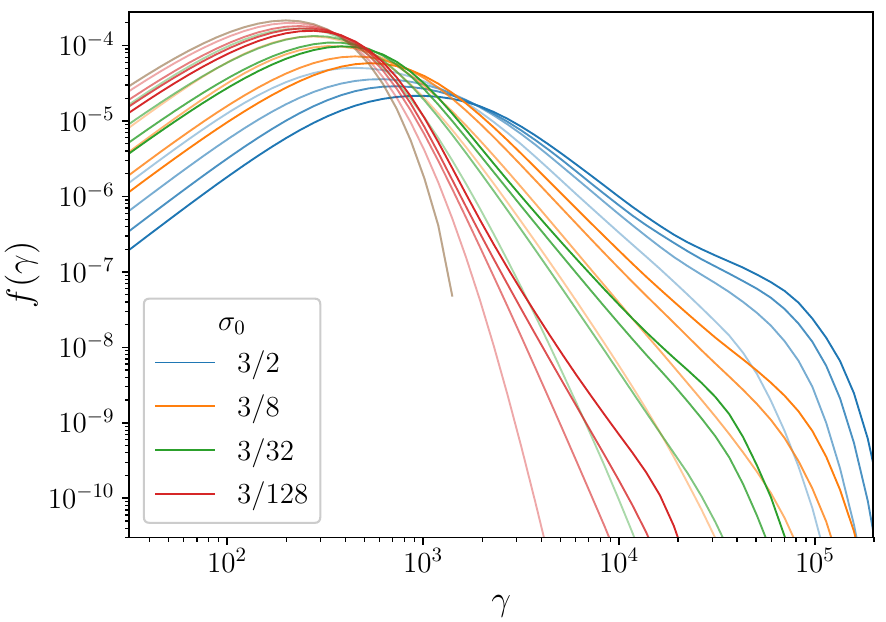}
\caption
{
Evolution of particle energy distribution for simulations with $N = 768$ and different $\sigma_0 \in~\{\sigmaFourVals\}$. Five times (separated by equal durations) are shown for each simulation, including the initial time, three intermediate times, and the final time (in order of decreasing color fade). 
}
\label{fig:evoEnergySpec}
\end{figure}

\autoref{fig:evoEnergySpec} shows the evolution of the overall particle energy spectrum $f(\gamma, t)$ for simulations with $\sigma_0 \in \{\sigmaFourVals\}$ and $N = 768$, which have the following common behaviour.
Beginning at a relativistic thermal distribution, each spectrum develops a power-law nonthermal tail after a few \Alfven{} crossing times.
This nonthermal segment begins near the average particle energy and extends to roughly the system-size limit $\gammaMax \equiv L e B_0 / 2 \restEnergy = L\gammaAvg/2\reo$, where the gyroradius reaches the system size.
For example, our typical runs with $\relLar = 512$ have $\gammaMax \simeq \num{7.7e4}$, while our largest simulation has $\relLar = 1024$ and $\gamma_{\rm max}\simeq \num{1.5e5}$.
As the system receives continuous energy input from turbulent driving and there is no radiative cooling, the plasma heats over time and so the extent of the nonthermal power-law segment shrinks as the simulation progresses. 
As particles are accelerated to the system-size limit, they form a pileup population near~$\gammaMax$, appearing as a bump at the end of the distribution \citep{zhdankin2018SystemsizeConvergenceNonthermal}.

\begin{figure}
\centering
\includegraphics[width=\linewidth]{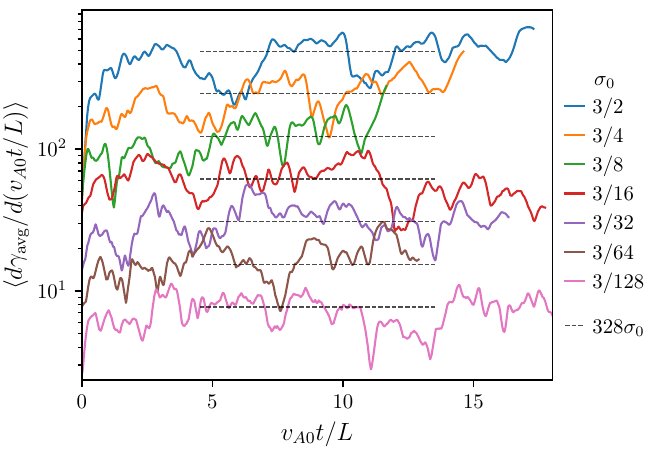}
\caption
{
Moving average of rate of change of average particle energy $\langle d\gamma_{\rm avg}/dt \rangle$ (normalized by $v_{A0}/L$) versus time for simulations with $N = 768$ and $\sigma_0 \in \{\sigmaScanVals\}$.
The width of the moving average is $L/\vao$, and this filtering is denoted by angled brackets.
Black dashed lines correspond to $\dot{\gamma}_{\rm avg} \propto \sigma_0$.
}
\label{fig:avgEnergy}
\end{figure}

\autoref{fig:avgEnergy} shows the plasma heating history, --- i.e., the time evolution of the rate of change of the global average particle energy~$\dot{\gamma}_{\rm avg}$, after applying a central moving time-averaging of width $L/\vao$ (based on the initial \Alfven{} velocity~$\vao$), for the $N = 768$ series of simulations with $\sigma_0 \in \{\sigmaScanVals\}$.
We observe that, aside from order-unity fluctuations, $\gammaAvg$ predominantly increases linearly with time.
There is an initial segment of several \Alfven{} crossing times with a slower rate of increase, corresponding to the period when some of the energy injected by driving goes towards establishing turbulent electromagnetic fields.
Once turbulence is fully established, these fields become statistically quasi-steady, and thereafter the supplied energy is predominantly converted into the plasma internal energy (heat and nonthermal particle acceleration), corresponding to a long-term linear increase in $\gammaAvg$ with time.
The roughly even separation of the $\dot{\gamma}_{\rm avg}$ lines in \autoref{fig:avgEnergy} corresponding to simulations spaced by fixed multiples in $\sigma_0$ imply that the rate of energy gain is roughly proportional to~$\sigma_0$.
To aid visual comparison, lines with constant $\dot{\gamma}_{\rm avg} \propto \sigma_0$ are overlaid on \autoref{fig:avgEnergy}.

\begin{figure}
\centering
\includegraphics[width=\linewidth]{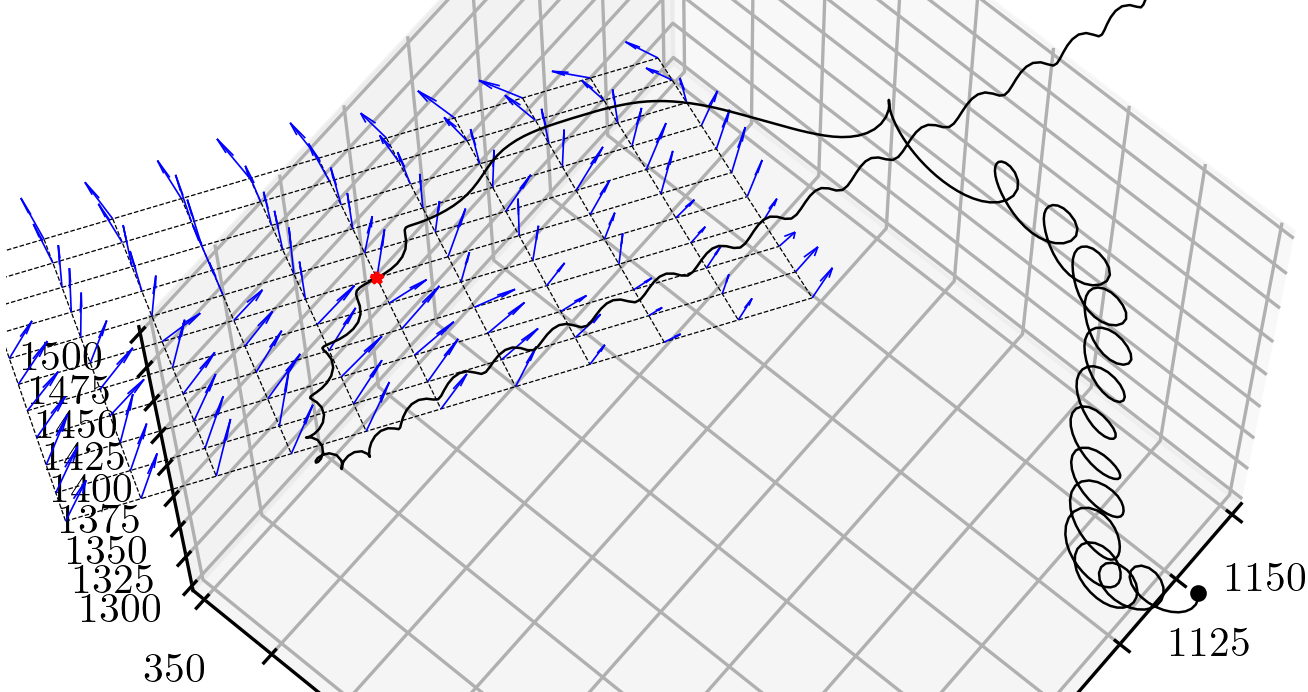}
\caption
{
Particle trajectory in the vicinity of an acceleration event, overlaid with magnetic field (blue vectors) at the time when the particle was located at the red marker. The latest time in the trajectory is marked by the black dot.
}
\label{fig:accFields}
\end{figure}

\autoref{fig:accFields} shows the particle  trajectory before, during, and after an  acceleration event, for a single particle in the simulation with $N =\nobreak 1536$ ($L/\reo=1024$) and $\sigma_0 = 3/8$ (selected arbitrarily from the tracked particle sample). The local magnetic field is also shown (blue arrows), taken at the time of the primary acceleration event (red dot). Before this energisation event, the particle enters the acceleration region while gyrating around a magnetic field line. It then encounters a shear in the magnetic field, which causes the gyro-centre motion to reflect, in a manner akin to the Fermi process \citep[][]{fermi_1949}, as considered in recent models such as \citet[][]{lemoine_2021, lemoine_2022, lemoine_2025}. The particle then leaves the region with a significantly larger Larmor radius than it began with, indicating an increase in the momentum component perpendicular to the magnetic field. While this is only a single example, we expect acceleration events such as this one to be generic within relativistic plasma turbulence. 

\section{Particle energy spectra} \label{sec:alpha}
Connecting the evolution of the particle energy distribution with the measured Fokker-Planck coefficients is important to better our understanding of turbulent particle acceleration.
If the system were steady-state, this would be simplified by the left-hand-side of \eqref{eqn:fpEne} being zero.
However, the time-dependence of $f(\gamma)$ described in \autoref{sec:global} complicates the interpretation of the Fokker-Planck coefficients in the context of a steady-state energy spectrum. Nevertheless, we still wish to understand the slowly evolving, quasi-steady state achieved at late times in these non-radiative turbulence simulations.
Understanding, even empirically, the key characteristics of~$f(\gamma,t)$, and in particular, the nonthermal tail, will help connect the time-dependent behaviour of the distribution function with that of the \ac{FP} coefficients measured in later sections.

A fundamental result of kinetic simulations of turbulent \ac{NTPA} is the generation of particle energy distributions with extended nonthermal ranges (see \autoref{fig:localPowerLawIndex}a).
For a particle energy distribution at some time~$t$, the local (in~$\gamma$) power-law index can be defined as $\alphaLocal(\gamma, t) \equiv \alphaDef$
(and hence a pure power-law segment is characterised by $\partial\alphaLocal/\partial\gamma = 0$). 
At a given instant, the local indices $\alphaLocal(\gamma)$ can be summarised into a representative single value, $\alphaSV(t)$, characterising the predominant value of $\alphaLocal(\gamma, t)$ in the high-energy nonthermal tail.
Moreover, a special value of~$\alphaSV(t)$, which we denote as~$\alphaFinal$, is often chosen to represent the late-time asymptotic (or ``final'') value of~$\alphaSV(t)$. 
These values $\alphaSV(t)$ and $\alphaFinal$ are main quantities of interest and have observational implications and theoretical significance.
For example, the system-size dependence of $\alphaFinal$ is critical to understand because it affects our ability to extrapolate the findings of these PIC simulations to astrophysical systems \citep{zhdankin2018SystemsizeConvergenceNonthermal}.
Therefore, in this Section, we investigate the time dependence of~$\alphaSV$, develop a prescription for measuring~$\alphaFinal$, and explore their dependence on system parameters such as the system size and initial magnetisation.

\subsection{Fitting the power-law index} 
\label{sec:fittingPlaw}
The process of obtaining $\alphaSV(t)$ from $\alphaLocal(\gamma,t)$ at a given time~$t$ is complicated by the necessity to select the section of $\alphaLocal(\gamma)$ to fit a straight line to.
This section must exclude, on the left, the thermal particles, and on the right, the high-energy pileup due to finite simulation size, all the while being as long as possible to maximise the accuracy of the fit.
Furthermore, one must then decide how to obtain the late-time asymptotic value~$\alphaFinal$.
For example, past studies have selected $\alphaFinal$ based on ``the time with the longest fitted power-law segment'' \citep{zhdankin2017KineticTurbulenceRelativistic} or a ``time that is logarithmic with system size'' \citep{zhdankin2018SystemsizeConvergenceNonthermal}.
These decisions limit the precision with which the evolution of different particle energy distributions can be compared, and thus a more systematic scheme is desirable.

Here, we describe a procedure to distill a unique $\alphaSV(t)$ value for a given distribution~$f(\gamma, t)$ that is well-suited to the particle energy distributions produced by our simulations and reasonably insensitive to fitting choices.
This scheme yields values similar to those obtained by the method developed by \citet{werner2018NonthermalParticleAcceleration}, but is more robust against the maximum-energy pileup at late times.

\begin{figure}
\centering
\includegraphics[width=\linewidth]{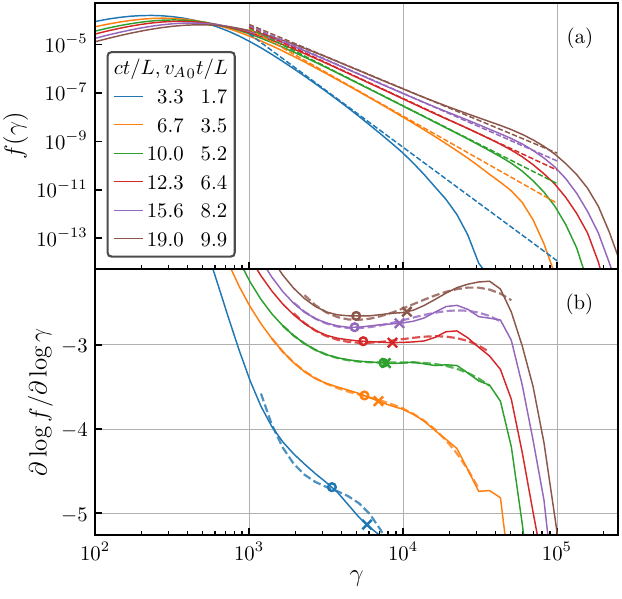}
\caption
{
Particle energy distribution $f(\gamma)$ (a) and the local power-law index $-\alpha_{\rm loc} = \partial \log{f}/\partial \log{\gamma}$ (b) at several different times for a representative simulation, having $\sigma_0 = 3/8$ and $L/\reo=512$. 
In (a), dashed lines show the power-law fits. In (b), dashed lines are the fitted cubics, crosses mark the geometric mean of $\gammaAvg$ and $\gammaMax$, while circles mark the first turning point of the cubic or the inflection point if there are no turning points. See the main text for more details.
}
\label{fig:localPowerLawIndex}
\end{figure}

At intermediate and late times, when the nonthermal tail has fully developed and a high-energy pileup has appeared, there is a local minimum in the local power-law index~$-\alphaLocal(\gamma)$ (\autoref{fig:localPowerLawIndex}b). 
Even as the thermal peak continues to move to the right and the maximum-energy pileup increases in size, the energy $\gamma$ at which this local minimum is attained remains very stable. Furthermore, precisely because this minimum is a local stationary point, the function $-\alphaLocal(\gamma)$ varies very little in its vicinity. That is, even if the minimum is not located exactly, the resulting difference in the selected value of the local index $\alphaSV$ is small.
This stability suggests considering the local minimum value of~$-\alphaLocal(\gamma, t)$ as the overall (time-dependent) power-law index~$\alpha(t)$ characterising the nonthermal part of the spectrum.
We therefore construct an automatic procedure to extract the local minimum of~$-\alphaLocal(\gamma,t)$.  We expect this procedure to be broadly applicable to other simulation studies which produce relatively clean and smooth particle distributions with similar structure, exhibiting a thermal bulk followed by a nonthermal power-law tail and finally a pileup and cutoff at highest energies.

Our procedure works as follows. We locate the local minimum by fitting a cubic, in $\log \gamma$ space, (\autoref{fig:localPowerLawIndex}b, dashed lines) to the $\gamma$-range within a factor of five of the geometric mean of the average energy $\gammaAvg$ and the system-size limited energy $\gammaMax$ (\autoref{fig:localPowerLawIndex}b, crosses).
The Lorentz factor corresponding to the desired local minimum (\autoref{fig:localPowerLawIndex}b, circles) is then the first turning point of the cubic, or the inflection point if there are no turning points. 
The cubed term is forced to be negative so that the first turning point is a local minimum (if it exists); points where the corresponding energy spectrum density $f(\gamma)$ is less than \num{e-9} of the peak value at that time are excluded due to noise.
We expect that other methods for finding local minima or inflection points would produce very similar results.
When the fit does not have a significant flat or inverted section, $\alphaSV$ is uncertain, but this simply reflects the fact that the power-law tail is not yet well developed at that time (e.g., \autoref{fig:localPowerLawIndex}, blue line).

This procedure does not have any physical or theoretical justification, and has so far only been applied to smooth spectra from simulations. However, it has some favourable properties.
As can be seen in \autoref{fig:localPowerLawIndex}a, it very closely fits the centre of the power-law range, and tends to trim off the late-time pileup at~$\gammaMax$.
Also, as it fits a turning point in $\alphaLocal \equiv \alphaDef$, it can be said to fit the straightest location of the power-law section of~$f(\gamma)$. 
The main idea of this procedure is to adopt the first local minimum value of $-\alphaLocal$ as the preferred single-value~$\alphaSV$ for a given energy spectrum; the exact method for locating this local minimum is secondary.

\subsection{Application to the system-size scan and statistical samples} \label{sec:alphaLscan}
\begin{figure}
\centering
\includegraphics[width=\linewidth]{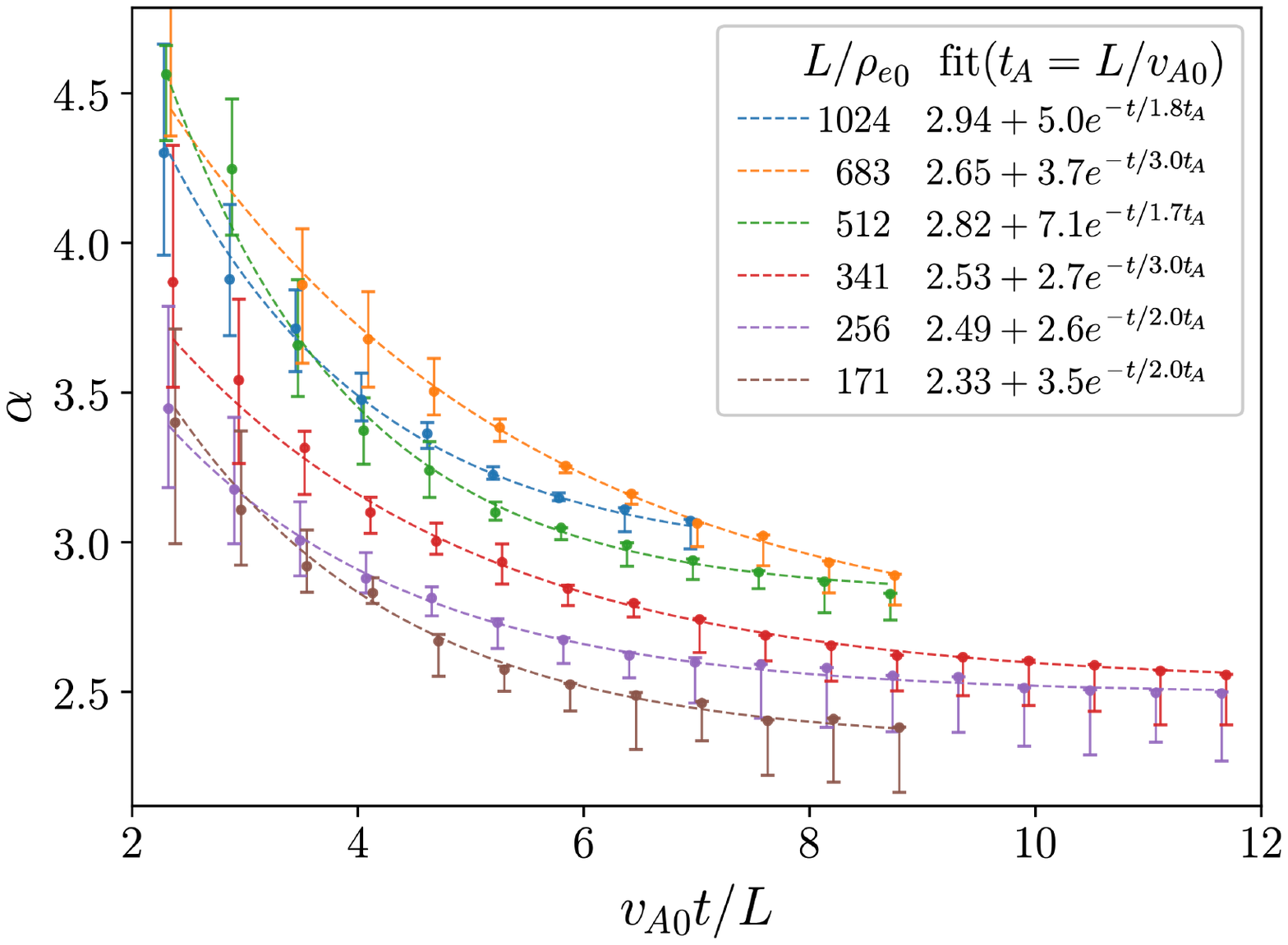}
\caption
{
Time evolution of power-law indices $\alpha$ for simulations with different system size $L/\reo$ but the same $\sigma_0 = 3/8$, with exponential fits (dashed lines).
Data points for each simulation are slightly time-shifted for visibility.
}
\label{fig:alphaExpL}
\end{figure}

We apply our procedure for identifying the single-value power-law index $\alphaSV$ on $f(\gamma, t)$-snapshots from the system-size scan simulations with fixed $\sigma_0 = 3/8$ and varying $L/\reo \in \{\relSizeVals\}$.
We ignore data points before $t = 2L/\vao$, when turbulence has not yet fully developed.
\autoref{fig:alphaExpL} shows the resulting~$\alphaSV(t)$.
The vertical line segments on each plotted point are obtained as follows: expand a symmetric $\alphaLocal$-range around $\alphaSV$ until it covers a surrounding contiguous $\gamma$-range of at least half a decade; the vertical lines then indicate the actual $\alphaLocal$-range included in this $\gamma$-range (neither of which must be symmetric).
These ``error bars'' indicate the range of surrounding $\alphaLocal$ values and should not be overly interpreted in a statistical sense. A strongly single-sided ``error bar'' indicates that $-\alphaSV$ lies in a local minimum of~$-\alphaLocal(\gamma)$.
We discard any data point with an ``error-bar'' range greater than 40\% of $\alphaSV$ as this indicates an insufficiently converged power-law segment.

%% Non scan-dependent
\autoref{fig:alphaExpL} shows that $\alphaSV(t)$ decays with time in every simulation.
Hence, we fit each $\alphaSV(t)$ with a three-parameter exponential decay:
\newcommand{\alphaExpEqn}{\alphaSV(t) = \alphaZero + \DeltaAlpha \exp(-t/\tauAlpha)}
\begin{equation}
\alphaExpEqn,
\label{eqn:expFit}
\end{equation}
and find that the fits are excellent.
Although having just warned against statistical interpretation of the ``error bars,'' we nevertheless inversely weigh the data points by their extent, as one would for true statistical standard deviations.

%% Not scan-dependent
The exponential fit clearly suggests $\alphaZero$ as a natural candidate for the late-time ``convergent'' value of~$\alphaSV$, i.e., for~$\alphaFinal$, with few arbitrary parameters.
The parameters $\DeltaAlpha$ and $\tau$ should be considered with respect to the convergence of the power-law tail, separately from its initial formation. Thus, $\DeltaAlpha$ is the degree to which $\alphaSV$ evolves over the convergence timescale~$\tau$.
By far the most robust parameter is~$\alphaZero$, which is essentially unchanged for any reasonable variant of the fitting parameters, whereas $\DeltaAlpha$ and $\tau$ are somewhat sensitive to the fit starting time.
The fits are reasonably insensitive to the maximum fitted time as long as the total duration is longer than about~$8L/\vao$. Hence, it is reasonable to expect that the trends fitted to the entire simulation as-is would continue if the simulation were somewhat longer. However, one would not expect it to hold indefinitely, as the nonthermal power-law range would eventually be squeezed from both sides until it disappears entirely.

\begin{figure}
\centering
\includegraphics[width=\linewidth]{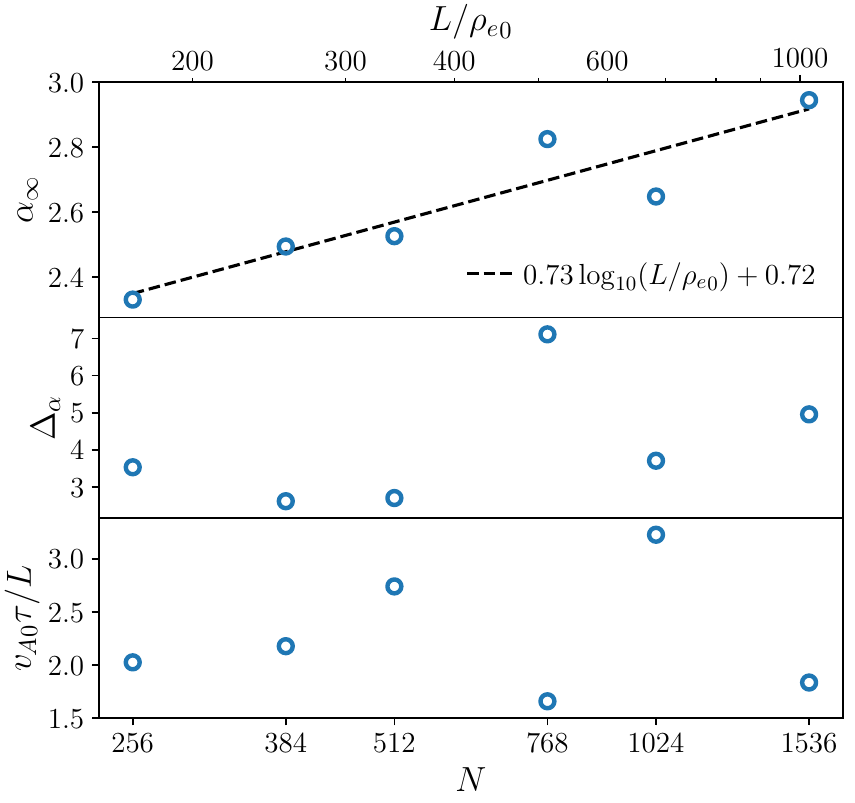}
\caption
{
Dependence of the three exponential fit parameters ($\alpha_\infty$, $\Delta_\alpha$, and $ \tau$ normalized to $L/v_{A0}$) of \eqref{eqn:expFit} on the relative system size~$L/\reo$ for simulations with $\sigma_0 = 3/8$. The fit by \eqref{eqn:a0fitL} is shown in a dashed line.
}
\label{fig:3pL}
\end{figure}

%% 3pL discussion
We use the fitted exponential parameters to quantify and compare the per-simulation time-evolution characteristics as functions of the relative system size~$L/\reo$.
\autoref{fig:3pL} shows that as $L/\reo$ increases through a factor of six, $\alphaZero$ increases modestly, while $\DeltaAlpha$ and $\tau$ have no particular trend.
This matches the basic observation in \autoref{fig:alphaExpL} that
$\alphaSV(t)$ tends to shift upwards in magnitude with increasing $L/\reo$, but otherwise behaves similarly between simulations.
The convergence timescale $\tauAlpha$ being a few \Alfven{} times matches the natural expectation for the system to evolve on \Alfven{ic} timescales.
The $\alphaZero$ data suggests that the simulations are not quite converged at the lower $L/\reo$ values, but there is not enough information to decide either way for the higher explored~$L/\reo$.
A reasonable fit for $\alphaZero$ as a function of $L/\reo$ is
\begin{equation}
\alphaZero = 0.73 \logten(L/\reo) + 0.72,
\label{eqn:a0fitL}
\end{equation}
and this is displayed in \autoref{fig:3pL}. 
The tendency for larger system size to result in steeper power laws may be attributed to the longer time (in terms of large-scale crossing times) required for particles to reach the system-size-limited energy \citep{zhdankin2018SystemsizeConvergenceNonthermal}. 

%% Stat sample
To get an idea of the statistical variability of the results, we analyse two sets of computationally cheaper simulations where the same system parameters were used multiple times with different random seeds.
One set used 7 simulations with $\sigma_0 = 3/8$, $\Nppc = 32$, and $N = 768$ $(L/\reo = 512)$; the other, consisting of 16 simulations, differs by using $N = 384$ $(L/\reo = 256)$.
The results are presented in ~\autoref{table:statSample}.
The scatter in $\alphaZero$ is only a few percent, whereas the variation levels of $\DeltaAlpha$ and $\tauAlpha$ are both substantial.
Therefore, it is plausible that the $\alphaZero$-trend observed in \autoref{fig:3pL} is real, but the variations in $\DeltaAlpha$ and $\tauAlpha$ are noise.

\begin{table}
\caption{Statistical variation in $\alpha(t)$ exponential fit parameters for sets of repeated simulations with the same physical parameters, listed as mean$\pm$standard deviation.
Both sets had $\sigma_0 = 3/8$ and $\Nppc = 32$.}
\label{table:statSample}
\begin{tabular}{ccccc}
    \hline
    $L/\reo$ & count & $\alphaZero$ & $\DeltaAlpha$ & $\vao \tauAlpha / L$ \\
    \hline
    512 & 7  & $3.00\pm0.06$ & $8.9\pm4$ & $1.3\pm0.2$ \\
    256 & 16 & $2.68\pm0.05$ & $7.6\pm4$ & $1.4\pm0.4$ \\
    \hline
\end{tabular}
\end{table}

\subsection{Application to the $\sigma_0$ scan}
\begin{figure}
\centering
\includegraphics[width=\linewidth]{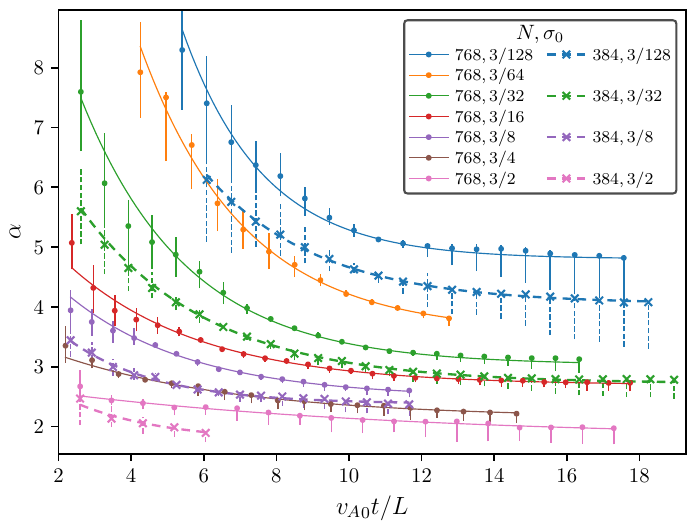}
\caption
{
The time evolution of the power-law index $\alpha$ for simulations with varying $\sigma_0$ and $L/\reo$, each fit by an exponential of the form \eqref{eqn:expFit} (solid and dashed lines). 
\autoref{fig:3pSig} presents the best-fit parameters.
}
\label{fig:alphaExpSig}
\end{figure}

\begin{figure}
\centering
\includegraphics[width=\linewidth]{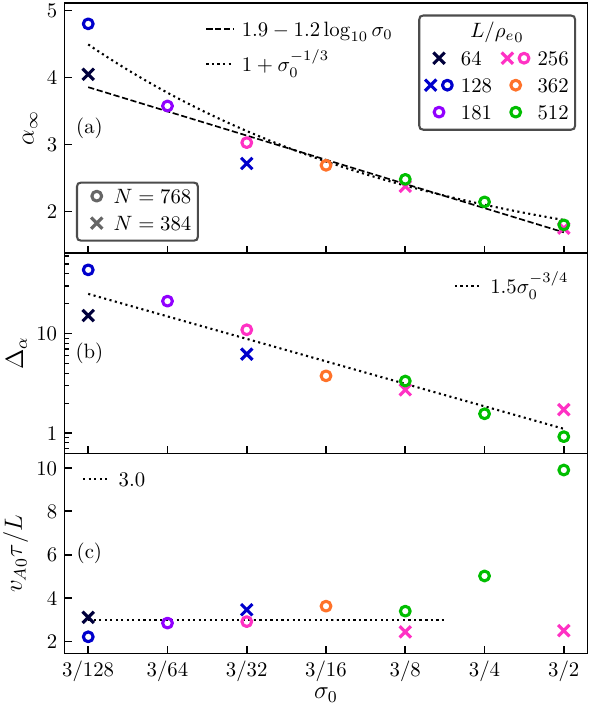}
\caption
{
Dependence of the three exponential fit parameters ($\alpha_\infty$, $\Delta_\alpha$, and $ \tau$ normalized to $L/v_{A0}$) of \eqref{eqn:expFit} on $\sigma_0$. Data points corresponding to different values of $L/\reo$ are indicated by different colors. Fits discussed in the text are shown in dotted or dashed lines.
}
\label{fig:3pSig}
\end{figure}

We repeat the above analysis on the \medSigScan{} and the \smlSigScan{} and combine the results.
\autoref{fig:alphaExpSig} shows the $\alphaSV(t)$ with exponential fits while \autoref{fig:3pSig} shows the exponential fit parameters versus~$\sigma_0$, grouped by~$L/\reo$.
As before, the exponential fits are good in all cases.
As $\sigma_0$ increases, $\alphaZero$ and $\DeltaAlpha$ decrease, while $\tauAlpha$ forms a roughly constant cluster around~$3L/\vao$ but displays a large scatter for $\sigma_0 \geq 3/4$.
These high-$\sigma_0$ values of $\tauAlpha$ should not be given too much credence. Indeed, one can observe in \autoref{fig:alphaExpSig} that ($N = 384$, $\sigma_0 = 3/2$) has few points, while ($N = 768$, $\sigma_0 = 3/2$) covers only a narrow range of $\alpha$-values, leading to a curve too shallow to reliably characterise the exponential.
Furthermore, \autoref{table:statSample} shows that $\tauAlpha$ already varies substantially for $\sigma_0 = 3/8$ and one expects even more relative variation for a shallower curve.
Overall, the illustrated trends in the fit parameters reflect the harder power-law distributions and faster system dynamics (with respect to~$L/c$) obtained at higher~$\sigma_0$, as previously noted by \citet{zhdankin2017KineticTurbulenceRelativistic}.

Even after combining the two $\sigma_0$ scans, there is not enough parameter variation ``orthogonal'' to $\sigma_0$ to properly characterise the effect of $L/\reo$ with this data set.
However, we note that for the four $\sigma_0$ values for which there are both $N = 768$ and $N = 384$ simulations,
$\alphaZero$ is slightly lower in the $N = 384$ case, which is consistent with the \autoref{sec:alphaLscan} findings.
\autoref{fig:alphaExpSig} also shows this relationship clearly, with $\alphaSV(t)$ tending to be lower in magnitude for $N = 384$ simulations compared to their $N = 768$ counterparts.

%% Fits vs sigma
To quantify the dependencies of $\alphaZero$ and $\DeltaAlpha$ on the system parameters observed in our simulations, we offer the following simple and practical empirical fits for these functions.
Firstly, a test-fit to $\alphaZero$ of a log-bilinear function of $\sigma_0$ and~$L/\reo$ indicated that the addition of~$L/\reo$ was statistically insignificant.
This is due to multicollinearity in the parameter space, not lack of effect, as \autoref{sec:alphaLscan} shows a significant trend for $\alphaZero$ against~$L/\reo$.
Consequently, we proceed with fits against just~$\sigma_0$, noting also that partitioning the data by~$L/\reo$ or~$L/\deo$ would not give substantially different results.

We find that the best log-linear fit of $\alphaZero$ as a function of $\sigma_0$ is: 
\begin{equation}
% \alphaZero \approx 2.0 - 0.5 \ln\sigma_0 \, .
\alphaZero \approx 1.9 - 1.2 \logten\sigma_0 \, .
\end{equation}
Alternatively, the offset power-law 
\begin{equation}
\alphaZero \approx 1+ \sigma_0^{-1/3}
\end{equation}
is also a good fit (the motivation for the constant term being equal to~1 is that this is a physical lower limit for~$\alpha$).
These fits are displayed in \autoref{fig:3pSig}a.

We also find that 
\begin{equation}
\DeltaAlpha = 1.5\sigma_0^{-3/4} 
\end{equation}
is a good fit for~$\DeltaAlpha$, as shown in \autoref{fig:3pSig}b.

%% Hedging about simulation coverage
We stress that more simulations would be needed to fully map out the dependence of the fit parameters over the 2D $(\sigma_0, L/\reo)$ parameter space, and also to further quantify the random variation of repeated simulation runs with the same initial system parameters.
We leave these further simulations required for more comprehensive results to future work.

%\clearpage
\section{Tests of diffusive particle acceleration} \label{sec:fp}
\subsection{Test procedure} \label{sec:fpTests}
This Section details our procedure to measure the \ac{FP} coefficients and their subsequent use in testing whether an energy diffusion-advection equation~\eqref{eqn:fpEne} is appropriate to model particle acceleration in these turbulence simulations.
This was found to be true for a single simulation in our previous study \citet{wong2020FirstprinciplesDemonstrationDiffusiveadvective}, and we improve on it here with more comprehensive tests on a larger parameter space.

The tracked particles described in \autoref{subsec:trackOsc} are integral to this analysis. 
First, we remove the large $\ExB$-drift-induced particle energy oscillations --- which are incompatible with an energy diffusion model --- by transforming to the particle's $\ExB$-drift frame, as described in \autoref{subsec:trackOsc}.
Then, for each simulation, we select approximately 100 evenly-distributed time instants, and
at each instant, divide particles into energy bins with edges at 10\% intervals, at $\gamma = 10 (1.1^0, 1.1^1, 1.1^2, ...)$.
We refer to each bin by its originating time instant $t_0$ and its arithmetic bin centre energy~$\bcGamma$.
For the population of particles in each bin, we compute the following quantities as functions of elapsed time~$\changeT \equiv t - t_0$:
the energy standard deviation~$\stdev$, energy variance~$\varEnergy \equiv \stdev^2$,
mean energy~$\binMean$, and change in the mean energy $\Delta\binMean \equiv \binMean(t) - \binMean(t_0)$ [where $\binMean(t_0) \approx \bcGamma$]. 
Examining the evolution of the bin particle population and these moments is a standard technique for analysing systems for \ac{FP}-type behaviour and subsequently measuring the \ac{FP} coefficients (\citealt{siegert1998AnalysisDataSets}, \citealt{friedrich2000ExtractingModelEquations}).
It is these procedures that require tracked particles and that cannot be done by just using summary statistics such as the overall particle energy distribution.

For the classical Brownian-type diffusion of the \ac{FP} equation, the bin energy variance increases linearly with time in the limit of infinitesimal elapsed time: $\lim_{\changeT \to 0} \varEnergy \propto \Delta t$.
Accordingly, a primary contraindication to \pgls{FP} model would be if the energy trajectories exhibit anomalous diffusion instead, e.g., where the energy histories are dominated by rare but large jumps such that the energy variance intrinsically scales nonlinearly with elapsed time.
This has been suggested to be the case in plasma turbulence, for instance, in \citet{isliker2017FractionalTransportStrongly} and \citet{isliker2017ParticleAccelerationFractional}.
Distinguishing definitively between conventional and anomalous diffusion can, however, be difficult in practice,
especially if anomalous effects are marginal, or only manifest rarely, in isolated or intermittent situations.
There are several reasons for this, but the most relevant one here is that if the \ac{FP} coefficients vary with particle energy, this can cause a nonlinear variance scaling at finite (but arbitrarily short) elapsed time while retaining a classical Brownian diffusion-advection equation
(this can happen with relevant and non-pathological scalings; we will give some concrete examples shortly, see also \citealt{friedrich2002CommentIndispensableFinite}).
Therefore, we take the approach of a consistency check:
we measure the \ac{FP} coefficients assuming standard Brownian diffusion, and then check whether a numerical solution of the \ac{FP} equation using the measured coefficients reproduces the \ac{PIC} simulation's particle-energy distribution evolution both globally and for individual bins (\autoref{sec:fpEvo}).
If the agreement is good, we can be confident that even if anomalous diffusion is present, it does not substantially affect the particle acceleration process, at least from the point of view of overall distribution characteristics like power law tails
(keeping in mind that this is the astrophysical observable of primary interest).

To measure the \ac{FP} coefficients from the bin moments, we use the following idealised equations for the latter's time evolution, which are obtained from the \ac{FP} equation \eqref{eqn:fpEne}
with a $\delta(\gamma - \bcGamma)$ initial condition and in the infinitesimal elapsed time approximation:
\begin{align}
\Delta\binMean(\bcGamma, \changeT) &= [\pdx D |_{\bcGamma} + A(\bcGamma)] \changeT \equiv M(\bcGamma) \changeT
\label{eqn:meant} \\
\varEnergy(\bcGamma, \changeT) &= 2 D(\bcGamma) \changeT
.
\label{eqn:stdevt}
\end{align}
However, the smallest elapsed time for which the moments are reliable is on the order of a gyroperiod. This is because, although the $\ExB$-based procedure mentioned above removes the bulk of the oscillation, some residual small-amplitude oscillation remains, possibly due to other types of drifts.
We also use linear fits to the moments to reduce noise, rather than measuring at a single point, and so, for the reason just mentioned, we begin these fits after one gyroperiod $2 \pi \bcGamma \me c / e \Brms$ (with $\Brms$ sampled at~$t_0$).
The fit end-times need further consideration, because longer fits better suppress noise, but also amplify the finite-time effects which lead to nonlinear time dependence of the bin moments \citep{gottschall2008DefinitionHandlingDifferent}.

To limit the impact of the nonlinear moments evolution
from finite-time effects, we terminate the fit once $\relStdev = 0.3$ (such that $\varEnergy/\bcGamma^2 \approx\nobreak 0.1$) or $|\Delta\binMean| / \bcGamma = 0.1$.
This uses $\bcGamma$ as an intuitive reference for the bin being relatively thin and close to its original position,
or equivalently, as a heuristic for how rapidly the \ac{FP} coefficients vary in energy space. 
These limits work well for Fermi acceleration, where $D \propto \gamma^2$ (temporarily ignoring~$A$ for simplicity).
Here, the strong~$D(\gamma)$ scaling produces effective superdiffusion that indeed begins when $\stdev \sim \bcGamma$ (one may verify this by referring to the analytical solutions for the equivalent geometric Brownian motion).
On the other hand, consider the \OU{} process, where $D$ is constant with respect to~$\gamma$ while~$A$ has a negative slope and arbitrary intercept (we show in \autoref{sec:ccLocal} that this is a reasonable conceptual model in certain energy ranges).
In this process, the variance becomes subdiffusive after an elapsed time inversely proportional to the advection coefficient's slope and hence at no particular value of $\relStdev$ or~$\relMeanChange$.
However, there is little we can do about this without turning to a complicated iterative procedure, and so, since $D \propto \gamma^2$ is the most important case for this study, we keep the fit limits referenced to $\bcGamma$ as described above.
The consistency check should also help to indicate whether this fitting range is reasonable.

We must also consider the \ac{FP} coefficients' time dependence because these simulations are not true steady states. This is due to overall system heating described in \autoref{sec:global}. This heating occurs on the \Alfven{} time-scale, so we cap the fit time to $\changeT <\nobreak L/\vao$.
Finally, we enforce a minimum duration of $0.1L/c$, for the rare cases where the above limits
result in a zero or negligible fit duration.

We use the above fit-range prescription to fit $\Delta\binMean$ and $\varEnergy$ to linear functions of elapsed time, then read off $D$ and $M$ from the fitted slopes using \eqref{eqn:meant} and \eqref{eqn:stdevt}.
We repeat this for all bins to obtain $D$ and $M$ for every point in our grid of $(\bcGamma, t_0)$ values, and elide the subscripts to consider them plainly functions of energy and time: $D(\gamma, t)$ and~$M(\gamma,t)$.
Finally, we obtain the advection coefficient using $A = M - \pdx D$, according to \eqref{eqn:meant}.

The rest of this Section examines
the bin energy moments (\autoref{sec:moments}) and the consistency check results (\autoref{sec:fpEvo}), leaving detailed analysis of the \ac{FP} coefficients to
\autoref{sec:dCoeff} and \autoref{sec:aCoeff}.

\subsection{Bin energy moments} \label{sec:moments}
\begin{figure*} %% var(t)
\centering
\includegraphics[width=\linewidth]{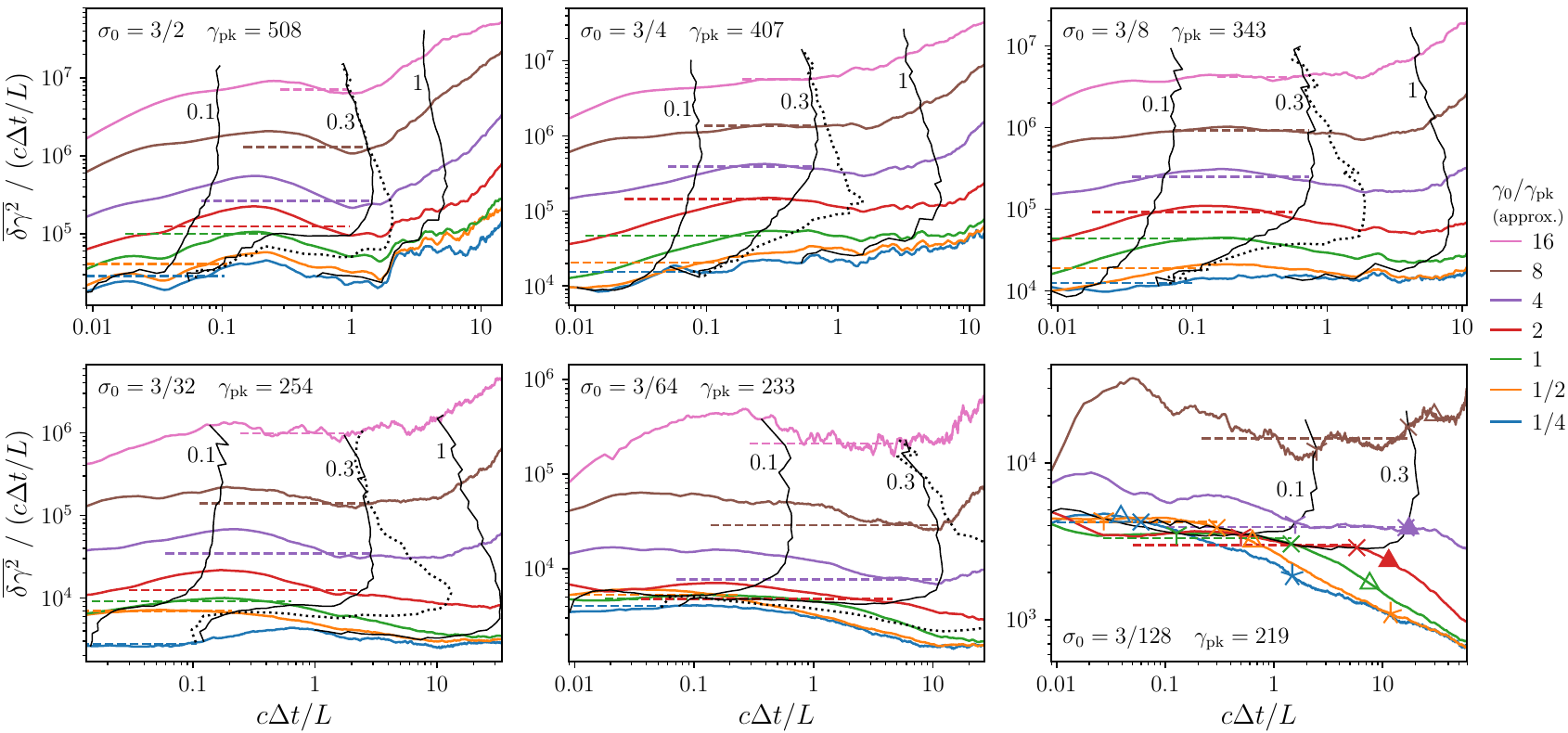}
\caption
{
Bin energy variance $\varEnergy(\changeT)$, compensated by $\cdtl$, where each panel corresponds to a simulation with different $\sigma_0$ (indicated on the top left of the panel). All simulations have $N = 768$ and the bins are initialised at $\binInitTime$.
Coloured dashed lines show the fits we use to obtain $D$; the line extent shows the fit range.
Black solid lines are contours of constant $\relStdev$ annotated by the corresponding $\relStdev$ value $\in \{0.1, 0.3, 1\}$.
Black dotted lines indicate $\relMeanChange = +0.1$.
These contours also use data from bins in between the ones we selected for plotting variance.
The $\sigma_0 = 3/128$ panel does not show the $\relMeanChange$ contour because $\relMeanChange$ is substantially negative for some bins in that simulation.
For that panel only, there are additional marks: for each variance line; an open triangle for $\relMeanChange = +0.1$; a filled triangle for $\relMeanChange = -0.1$; and three, four, and five armed spindles for $\relStdev = \{0.1, 0.3, 1\}$ respectively.
}
\label{fig:vart}
\end{figure*}

\autoref{fig:vart} shows the bin energy variance as a function of elapsed time, $\varEnergy(\changeT)$,
compensated by~$\cdtl$,
for a selection of simulations from the \medSigScan{}.
The bins shown are initialised at $\binInitTime$ and range from thermal ($\pkMult \approx 1$) to nonthermal ($\pkMult \gg \nobreak 1$) energies.
The plots show the linear fits that will be used to obtain $D$ via \eqref{eqn:stdevt}, which seem reasonable considering the heavy late-time bias inherent in a logarithmic scale.
They also show contours of constant $\relStdev$ and $\relMeanChange$ (see caption for details), which give a sense of the relative evolution timescales as a function of initial bin energy.

Generally, while $\relStdev \ll 1$, $\varEnergy$ increases roughly linearly with $\changeT$, suggestive of classical diffusion.
For low initial magnetisations $\sigma_0 \lesssim 3/32$, there is substantial energy subdiffusion at thermal energies $\bcGamma \lesssim \gammaPeak$, with power law indices $\mathbin{\sim} 0.7\textrm{--}0.9$ depending on~$\sigma_0$, which tends to become pronounced after intermediate times when $\relStdev > 0.3$.
\autoref{sec:ccLocal} shows that \ac{FP} evolution reproduces this.
There is also superdiffusion at late times, once $\stdev \sim \bcGamma$, which is particularly apparent at nonthermal energies and at higher~$\sigma_0$.
As mentioned in \autoref{sec:fpTests}, this would be consistent with (but not exclusive to) $\Dpropxx$.
It is possible to read off the rough diffusion-coefficient energy scaling from the contours of constant~$\relStdev$.
A vertical contour implies, through \eqref{eqn:stdevt}, that $\Dpropxx$; there are such near-vertical segments at high energies in all simulations.
Similarly, contours slanting from lower left to upper right, or from lower right to upper left, mean, respectively, a shallower or steeper scaling than~$\gamma^2$.

\begin{figure*}
%% mean(t)
\centering
\includegraphics[width=\linewidth]{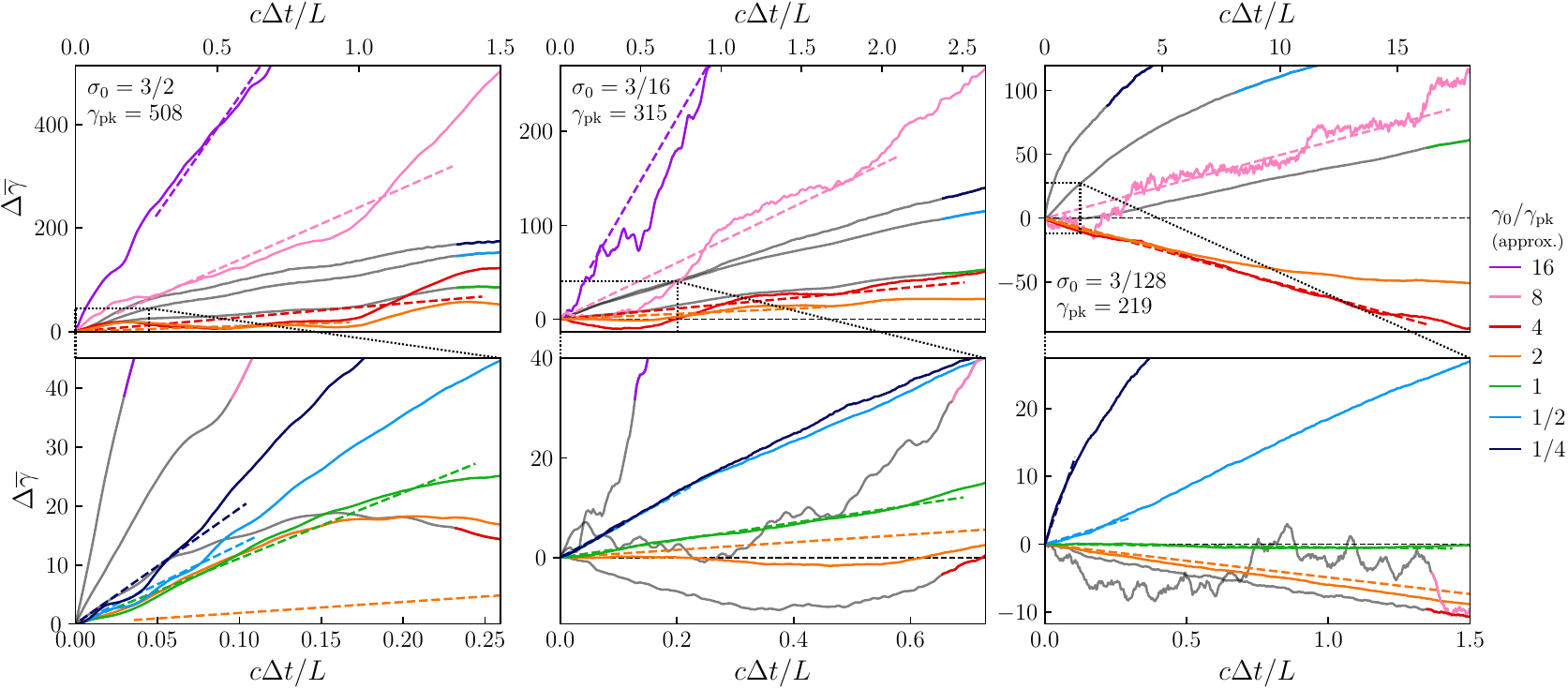}
\caption
{
Change in bin mean energy $\Delta\binMean(\changeT)$ for $\sigma_0 \in \{\sigmaThreeVals\}$ simulations (in order from left to right column), with $N = 768$ and bins initialised at $\binInitTime$.
The top row is best for viewing the lines for the high-energy bins while the bottom row presents the same data but zooms in on the black dotted box in the lower left corner, which is better suited for viewing the low-energy bins. 
In the top row, high-energy bin lines are fully coloured while only the tips of low-energy bin lines are coloured; this is reversed in the bottom row. 
}
\label{fig:meant}
\end{figure*}

\autoref{fig:meant} shows the change in mean bin energy as a function of elapsed time, $\Delta\binMean(\changeT)$.
The traces are generally linear.
The particle bin with minimum $\Delta\binMean(\changeT)$ slope for each given simulation (including negative slope as in the case of $\sigma_0 = 3/128$) tends to occur at central energies $\pkMult \approx 2\textrm{--}4$ rather than the lowest energies.
For the lowest $\sigma_0 =\nobreak 3/128$ simulation, the $\pkMult \leq\nobreak 1/2$ bins have flattening curves.
This occurs at long times and is likely due to coefficient energy dependence; we will check in detail for the $\pkMult =\nobreak 1/2$ bin in the next Subsection.

\subsection{Consistency checks}
\label{sec:fpEvo}
%% Subdiffusion consistency check
\label{sec:ccLocal}

\begin{figure}
\centering
\includegraphics[width=\linewidth]{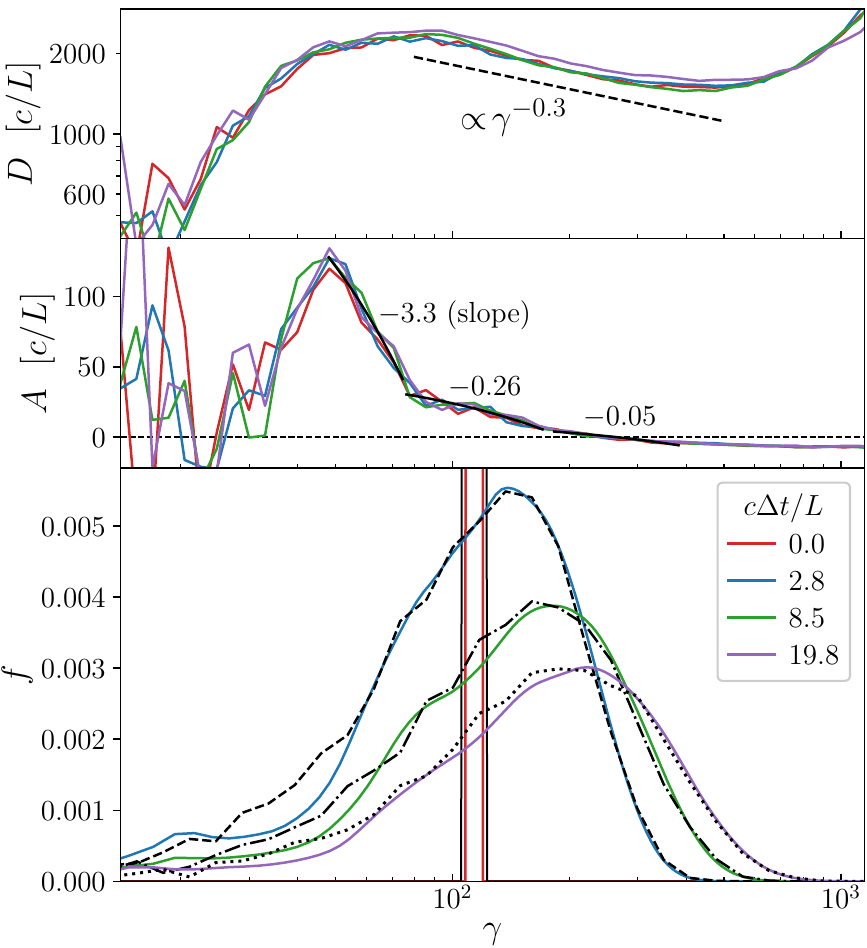}
\caption
{
Histogram of single-bin tracked particle energy (black lines in bottom panel) for four elapsed times $\Delta t \in \{0.0, 2.8, 8.5, 19.8\} L/c$, compared to the evolution of a matching initial narrow distribution using the \ac{FP} equation with measured coefficients (solid lines in bottom panel). There is an excellent match except at lowest energies, where $\pdx D$ is noisy and hence so is~$A$.
The local measured \ac{FP} coefficients are also shown (top and middle panel), with representative power-law fits (black lines): the negative slope of $A$ produces effective subdiffusion.
The bin is taken at $\binInitTime$ and has centre energy $\pkMult \approx 1/2$.
The simulation has $N \mathbin{=}\nobreak 768$ and $\sigma_0 \mathbin{=}\nobreak 3/128$.
The bin distribution function is normalised such that it integrates to~1.
}
\label{fig:singleBinVsx}
\end{figure}

\begin{figure}
\centering
\includegraphics[width=\linewidth]{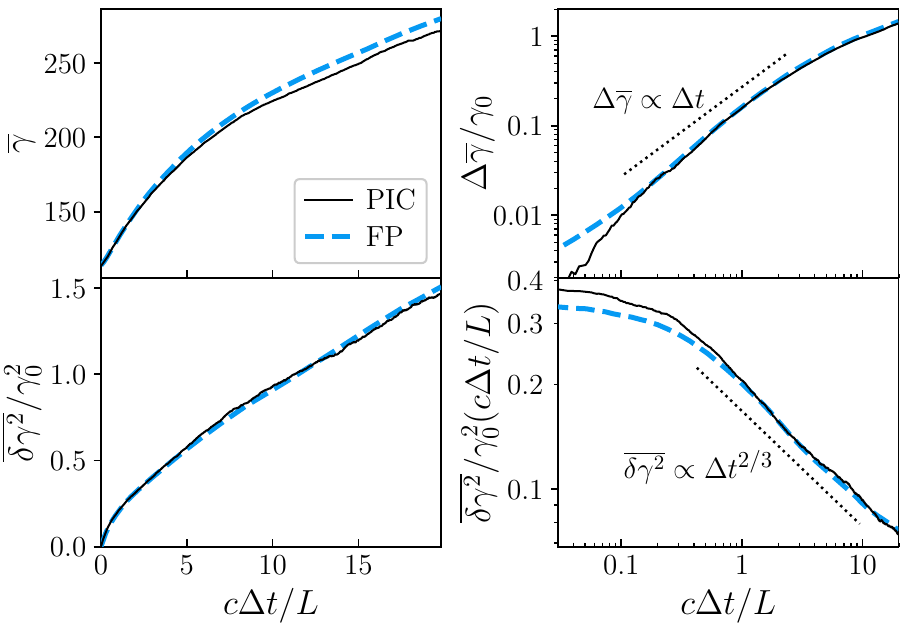}
\caption
{
Comparison of \ac{PIC} tracked particle and \ac{FP} evolution single-bin mean and variance versus time in linear (left) and log (right) scales. The log-scale variance panel is compensated by $\cdtl$. There is excellent agreement between the moments from the binned tracked particles (black solid lines) and those obtained by evolving a matching initial narrow distribution using the \ac{FP} equation with measured coefficients (blue dashed lines).
For large elapsed time, the variance evolves with a subdiffusive
scaling $\varEnergy \propto \changeT^{2/3}$ (dotted line).
The mean change does not have a definite power-law relationship with time ($\mathop{\propto}\changeT$ dotted line shown for reference).
The bin is taken at $\binInitTime$ and has centre energy $\pkMult \approx 1/2$.
The simulation has $N \mathbin{=}\nobreak 768$ and $\sigma_0 \mathbin{=}\nobreak 3/128$.
}
\label{fig:singleBinVst}
\end{figure}

In this Subsection, we test whether numerical \ac{FP} evolution with the measured coefficients reproduces the \ac{PIC} tracked particle distribution evolution, first for a single bin, and then for the entire particle energy distribution.

We choose a bin which shows subdiffusion, corresponding to the $\pkMult \approx 1/2$ line in the $\sigma_0 =\nobreak 3/128$ panel of \autoref{fig:vart}, and set the initial condition to a matching narrow rectangular distribution.
\autoref{fig:singleBinVsx} shows the subsequent energy distribution evolution along with the nearby \ac{FP} coefficient values, while \autoref{fig:singleBinVst} shows the bin energy moments as a function of time.
The excellent match suggests that any effective departure from classical diffusion is due to the \ac{FP} coefficients' energy dependence effected over finite elapsed time, and not anomalous diffusion.
We note that the coefficients are remarkably stable over this time period, as is consistent with the slow dynamics of low magnetisation simulations, and so coefficient time dependence is likely unimportant.

To get a better idea of what aspect of the \ac{FP} coefficients' energy dependence results in this subdiffusion, we compare to the \OU{} process mentioned in \autoref{sec:fpTests}. This should be a reasonable model because in the vicinity of this relatively low-energy bin, the diffusion coefficient has only a weak scaling of roughly $\gamma^{-0.3}$ while the advection coefficient has a negative slope.
For an advection coefficient slope of $-\OUslope$ (i.e., $A = \mathrm{const} -\OUslope \gamma$),
the \OU{} process elicits substantial variance subdiffusion after roughly $\changeT \sim |2\OUslope|^{-1}$.
At the chosen bin centre energy, $\OUslope \approx -0.3c/L$,
giving a nonlinear timescale of $\mathbin{\sim} 2L/c$, which is a reasonable match to when the variance in \autoref{fig:singleBinVst} becomes significantly nonlinear,
considering that this estimate heavily simplifies the \ac{FP} coefficients' energy dependence.
Furthermore, \autoref{fig:vart} shows that, up to at least $\pkMult = 2$, the subdiffusive departure becomes later as initial bin energy increases, which is consistent with \autoref{fig:singleBinVsx} showing that the magnitude of $\OUslope$ decreases with increasing energy.
These observations suggest that the observed subdiffusion is caused mainly by the negative slope in~$A$.

\begin{figure*}
\centering
\includegraphics[width=\linewidth]{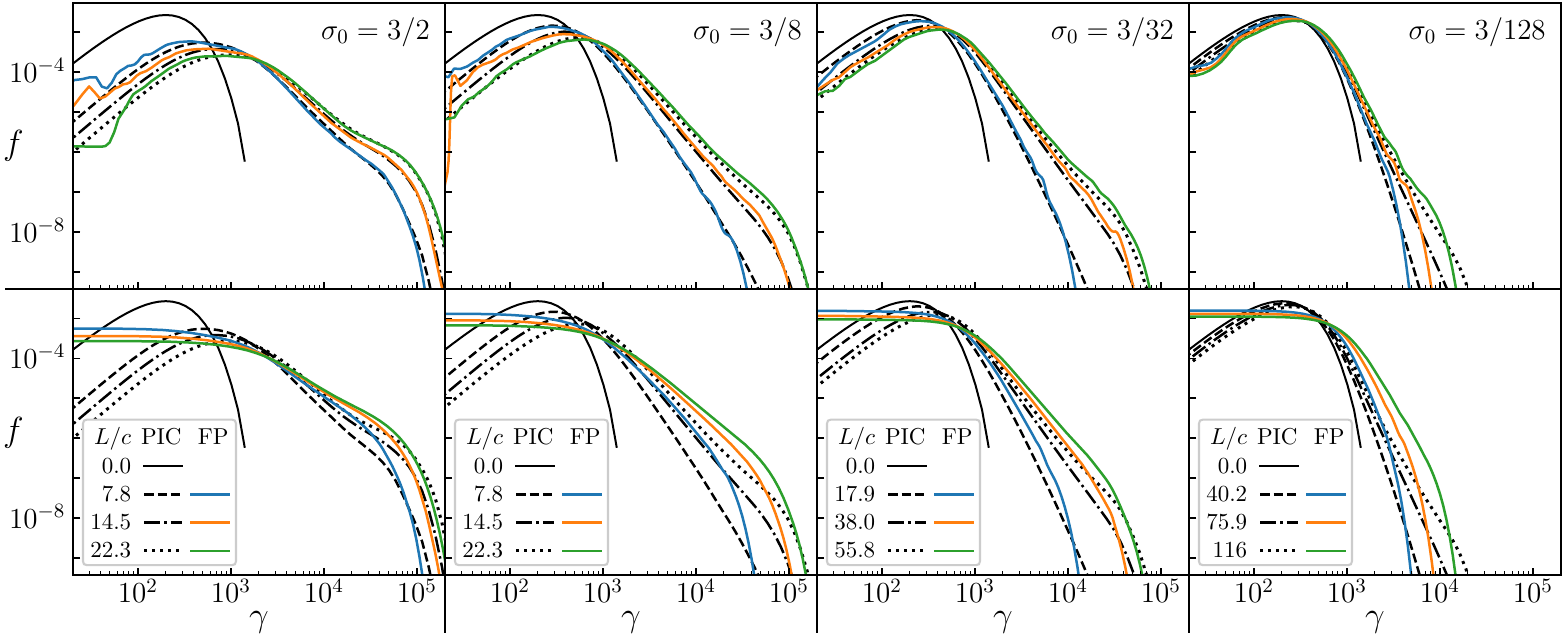}
\caption
{
Top row: global particle energy distribution $f(\gamma)$ for the PIC simulations at four times (black lines), compared to the numerical solution of the \ac{FP} equation (solid colored lines) using measured $D$ and $A$. The \ac{FP} evolution was initialised with the initial thermal distribution. Simulations with varying $\sigma_0 \in \{ 3/2, 3/8, 3/32, 3/128 \}$ are shown in the different columns, with excellent agreement between \ac{FP} and \ac{PIC} over a broad range of energy in most cases.
Bottom row: same as top row except with $A = 0$, which results in clear discrepancies between \ac{FP} and \ac{PIC}.
}
\label{fig:fpEvolution}
\end{figure*}

We now test whether the numerical solution of the \ac{FP} equation with these tracked-particle-based coefficients can reproduce the actual evolution of the full particle energy distribution observed in the PIC simulations.
\autoref{fig:fpEvolution} presents a comparison of the numerical FP and PIC evolution results for the simulations with $\sigma_0 \in \{\sigmaFourVals\}$ from the \medSigScan{}.
The initial condition is the initial thermal distribution. We find that the FP equation with the measured time-dependent coefficients accurately reproduces the evolution of $f(\gamma)$ from the \ac{PIC} simulations over the entire $\sigma_0$-range of simulations.
This provides solid evidence that the particle acceleration resulting in the nonthermal distribution can be completely modelled by the \ac{FP} equation~(\ref{eqn:fpEne}).

It is worth noting that, without the advection coefficient, the \ac{FP}-based distributions have a substantially altered shape. They still exhibit an evolving power-law tail at high energies, but with greater overall energy injection and generally harder power laws than the \ac{PIC} results. 
As was stated in \citet{wong2020FirstprinciplesDemonstrationDiffusiveadvective}, but here shown for a range of~$\sigma_0$, the advection coefficient therefore plays an important role in maintaining the shape of the particle energy distribution.

%\clearpage
\section{Diffusion coefficient} \label{sec:dCoeff}
Our previous study, \citet{wong2020FirstprinciplesDemonstrationDiffusiveadvective}, measured the energy diffusion coefficient $D$ in a single large simulation with moderate initial magnetisation, finding that $D$ scales as $\gamma^2$ in the nonthermal region and somewhat more shallowly at lower energies.
This Section characterises $D(\gamma,t)$ in simulations with different parameters; we examine the time and energy dependence of~$D$, and to what extent the aforementioned scaling characteristics persist.
Finding universal $\gamma^2$ scaling in the nonthermal range,
we extract the $\gamma^2$ prefactor $D_0$ and compare it to \ac{NTPA} theory predictions.

To investigate the effects of $\sigma_0$ and~$L/\reo$, we employ the simulations from the \medSigScan{} and the system-size scan.
We also use a smaller $N \speq 384$ $\Nppc$-scan to test convergence.
Please refer to \autoref{table:sims} and surrounding text for the complete parameter sets.

We can only measure $D$ for energy bins with enough tracked particles, which is dictated by the hardness and extent of the particle energy distribution at the measurement time.
To limit statistical noise, we discard results from bins with less than 10 particles.
Nevertheless, our results are primarily established by measurements at energies for which there are ample statistics, with thousands of tracked particles per bin.

\subsection{Magnetisation dependence} \label{subsec:Dsigscan}
% \subsection{The diffusion coefficient as a function of energy and time}
% \label{subsec:Dxt}
%% D(x,t)

\begin{figure*}
\centering
\includegraphics[width=\textwidth]{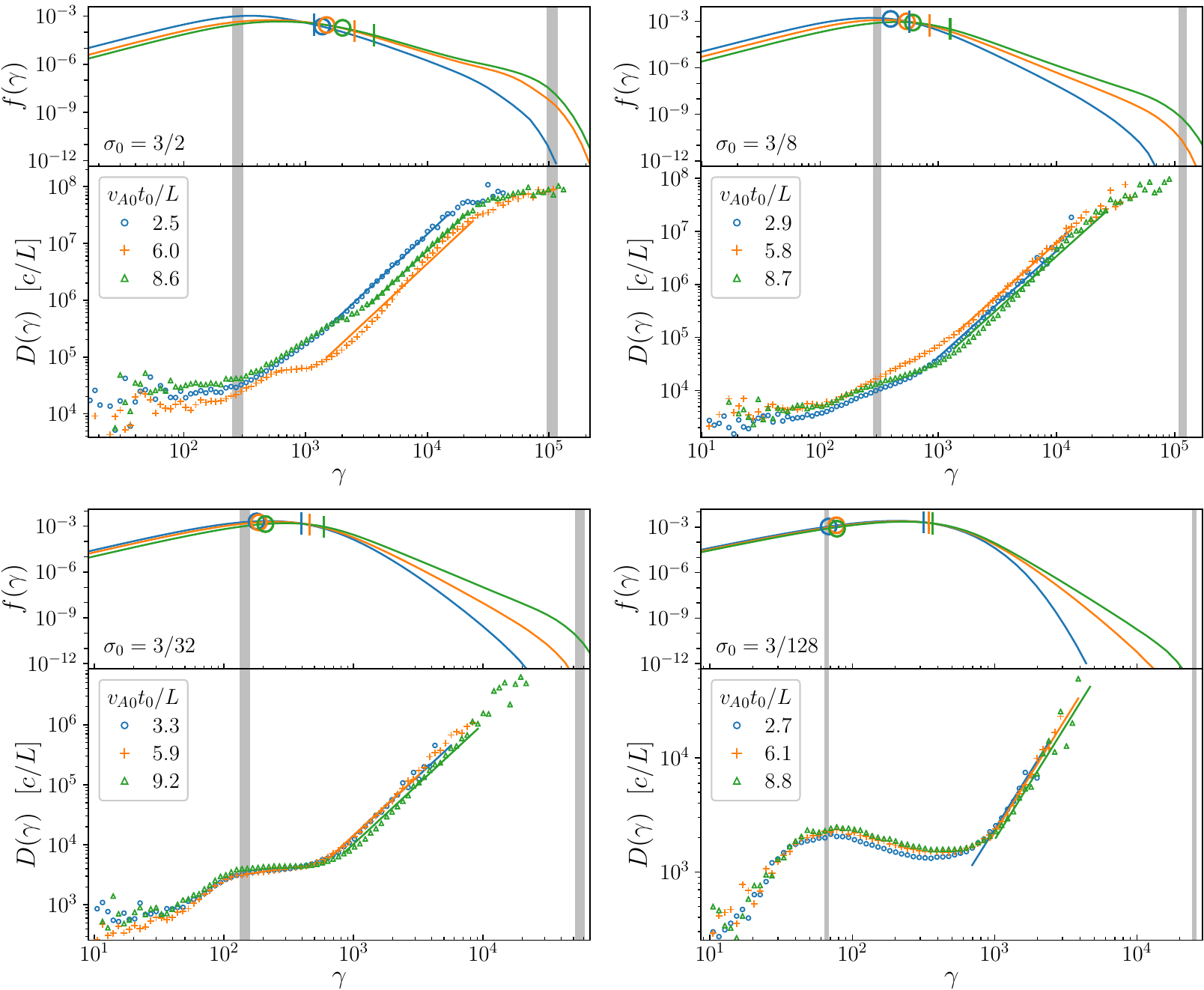}
\caption
{
The diffusion coefficient $D$ (lower subpanels) and particle energy distribution $f$ (upper subpanels) as a function of energy $\gamma$ for $\sigma_0 \in \{ 3/2, 3/8, 3/32, 3/128\}$ (as labeled in each panel) and three different times (indicated in legend), displaying the relevant plasma and simulation length scales (converted to particle energies via the Larmor radius) as well as the $D \sim \gamma^2$ fits to the nonthermal region that are used to extract~$D_0$ (solid coloured lines in lower subpanels).
The plasma length scales are the average Larmor radius, corresponding to the average particle energy $\rho_e$ (small vertical bars); and the skin-depth $d_e$ (open circles).
The simulation length scales are the grid resolution $\Delta x$ (vertical grey strip at lower energy); and the system-size limit $L/2$, corresponding to the high-energy cutoff $\gamma_{\rm max}$ (vertical grey strip at higher energy). For these respective quantities, the strips cover the 10th to 90th percentile values recorded between the times corresponding to the first and last plotted $t_0$ instance in each plot.
}
\label{fig:Dxt}
\end{figure*}

This Subsection uses the \medSigScan{} to analyse the dependence of $D(\gamma,t)$ on both the initial magnetisation $\sigma_0$ and the instantaneous magnetisation $\sigma(t)$.
\autoref{fig:Dxt} shows $f(\gamma,t)$ and $D(\gamma,t)$ for the simulations with $\sigma_0 \in \{\sigmaFourVals\}$ at times \figXtTimes{}.
These plots have several types of markers indicating different physical length-scales.
Each length is converted to the particle energy that would result in a gyroradius equal to that length by multiplying by $e \Brms / \me c^2$. The length-scales are:
\begin{enumerate}
    \item The grid resolution~$\Delta x$. At energies near and below the grid-resolution energy, data are likely to be affected by numerical noise.
    \item The system-size gyroradius limit~$L/2$, corresponding to the system-size energy limit $\gammaMax(t) \equiv L e \Brms / 2 \restEnergy$. This is close to the driving wavelength ($\Ld = L$ in our simulations) and so in the vicinity of $\gammaMax(t)$ we may see both finite system size and finite driving scale effects.
    \item The skin depth $d_e(t) \equiv \deEqn$.
    \item The characteristic gyroradius $\rho_e(t) \equiv \smash{\reEqn}$ (corresponding to $\gammaAvg$ by definition).
\end{enumerate}

We observe that $D(\gamma)$ can be divided into two segments by the beginning of the $f(\gamma)$ nonthermal power law.
In the nonthermal power law, we find a universal scaling $D \propto \gamma^2$ for all $\sigma_0$ (flattening slightly at highest energies, perhaps due to encroaching on~$\gammaMax$).
The beginning of this $D \propto \gamma^2$ appears to coincide with the start of the $f(\gamma)$ nonthermal power law, rather than other energy scales such as $\gammaAvg$, as evidenced by the $\sigma_0 = 3/128$ plot of \autoref{fig:Dxt}, where $\gammaAvg$ occurs at a substantially lower energy than the start of the $f(\gamma)$ nonthermal power law.
The low-energy scaling of $D(\gamma)$ depends on~$\sigma_0$.
For higher $\sigma_0 \gtrsim 3/8$, the low-energy scaling is consistent with a power law shallower than~$\gamma^2$.
For lower $\sigma_0 \lesssim 3/32$, there is an intermediate energy region with flatter scaling (see $\sigma_0 = 3/32$) or inverted scaling (see $\sigma_0 = 3/128$), and then a steeper section at lowest energies.

%% Low energy range
A natural explanation for the existence of the intermediate energy region in lower-$\sigma_0$ simulations is that certain length-energy scales are substantially separated only in those simulations.
However, near each end of the intermediate region there are multiple physical energy scales, making it difficult to unambiguously identify which of them are relevant.
For instance, some plausible delineators of the intermediate region are: $d_e$ and $\rho_e$, the grid scale and the larger of $d_e$ and $\rho_e$, and the grid scale and the beginning of the nonthermal power law.
A more thorough analysis with a targeted series of simulations would be needed to disentangle the effects of the different length-energy scales on~$D(\gamma)$.
We leave such work to future studies.

%% D0
Putting aside the low-energy behaviour, we now focus on the prefactor of the nonthermal scaling $D_0 \equiv D(\gamma) / \gamma^2$, its time evolution, and its dependence on $\sigma_0$ and~$\sigma(t)$.
The value of $D_0$ and its relation to other system parameters is important for testing various particle-acceleration theories, including the ones described in \autoref{sec:analyticalModel}.
In particular, standard second-order Fermi acceleration theories (see \autoref{sec:FPeq}) yield $D_0 \propto \uA^2$, where
$u_A \equiv \vA (1 - \vA^2/c^2)^{-1/2}$, which becomes $D_0 \propto \vA^2 \propto \sigma$ in the nonrelativistic turbulence limit $\vA \ll c$. 
However, a more sophisticated modern analysis by \citet{demidem2020ParticleAccelerationRelativistic}, also based on quasilinear theory but accounting for resonance broadening of  Alfv\'{e}n modes, yields 
$D_0 \propto \vA^3 \sim \sigma^{3/2}$ for Alfv\'enic turbulent cascades, while still retaining the conventional $D_0 \propto \vA^2 \propto \sigma$ scaling prediction for compressional fast and slow magnetosonic cascades. 
Furthermore, since our simulations are inhomogeneous and time-dependent, $\vA/c \equiv [\sigma/(\sigma + 1)]^{1/2}$ is only a characteristic definition; we do not know \emph{a priori} whether to emplace $\sigma_0$ or $\sigma(t)$ here for the purpose of comparison with Fermi acceleration models.
A careful analysis of first-principles PIC simulation results can help resolve this important issue. 
Finally, $D_0$ is also important because $D_{\smash{0}}^{-1}$ has a simple intuitive interpretation as the diffusive acceleration time for a particle in the nonthermal energy range; its energy doubles in a time of order~$D_{\smash{0}}^{-1}$, absent other effects (which can be seen by inserting $D = D_0 \gamma^2$ into \autoref{eqn:meant}).

%% D0(t) and sigma(t)
\begin{figure}
\centering
% \includegraphics[width=\linewidth]{20230606_p2dsg_d0tsigtSigScan.pdf}
% \begin{overpic}[width=\linewidth]{20230606_p2dsg_d0tsigtSigScan.pdf}
\begin{overpic}[width=\linewidth]{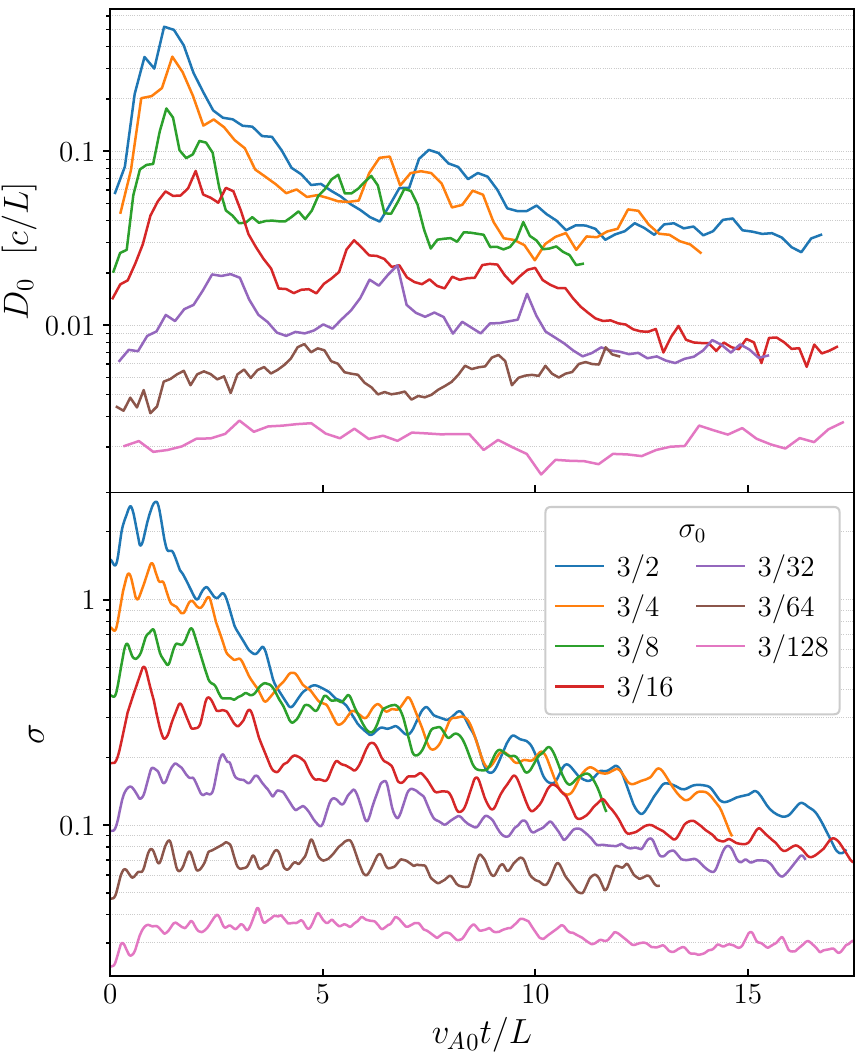}
    \put (40, 95) {(a)}
    \put (40, 49) {(b)}
\end{overpic}
\caption
{
(a) The diffusion coefficient prefactor $D_0(t)$, and
(b) the instantaneous magnetisation $\sigma(t)$,
for the \medSigScan{}.
The $D_0(t)$ and $\sigma(t)$ curves are qualitatively similar. Both show an initial increase followed by decay, with higher $\sigma$ also corresponding to higher $D_0$.
}
\label{fig:d0tsigtSigScan}
\end{figure}

We measure $D_0$ as a fine-grained function of time by fitting $D =\nobreak D_0 \gamma^2$ to the high-energy region in each sample of $D(\gamma, t)$.
\autoref{fig:d0tsigtSigScan} presents $D_0(t)$ and $\sigma(t)$ for the \medSigScan{}.
The resemblance between the two panels suggests a direct relationship between $D_0$ and $\sigma$, in accordance with theoretical predictions.
Nevertheless, let us first examine them separately.

The $\sigma(t)$ trajectories show a characteristic pattern of an initial increase to roughly twice $\sigma_0$ followed by decay.
The initial increase is caused by $\delta B$ going from zero to roughly $B_0$ as the turbulence is established, thus doubling the total magnetic energy density (as described in \autoref{sec:sims}, $\delta B \sim B_0$ is by choice).
Thereafter, as the plasma heats over time in our simulations, the relativistic enthalpy density increases, and this causes the magnetisation (being inversely proportional to the enthalpy) to decline.
We refer to section~4.2 of \citet{zhdankin2018NumericalInvestigationKinetic} for further details about $\sigma(t)$ time dynamics. 
The early-time decay in $\sigma$ is much more rapid for higher-$\sigma_0$ simulations; this is because the ratio of dissipated magnetic energy (per dynamical time) to initial thermal energy increases in proportion to~$\sigma$, as described quantitatively in section~4.2 of \citet{zhdankin2018NumericalInvestigationKinetic}. 
As a consequence of this $\sigma_0$-dependent decline of~$
\sigma$, simulations beginning at $\sigma_0 \geq 3/8$ reach similar values of~$\sigma$ at late~$\tva$.
Moreover, in absolute time units $ct/L$, the highest $\sigma_0$ simulations (e.g., $\sigma_0 = 3/2$) arrive at lower $\sigma(t)$ values faster than slightly lower $\sigma_0$ simulations (e.g., $\sigma_0 = 3/4$).

As alluded to earlier, the time evolution of $D_0$ is similar to that of~$\sigma$: initially, $D_0$ increases briefly, and thereafter it tends to decline.
This initial increase lasts for around one to three \Alfven{} crossing times.
The subsequent decline reflects the overall heating trend with both $D(\gamma)$ and $f(\gamma)$ moving to the right as the simulation progresses.
However, there are significant and sustained ($\sim\!\!L/\vao$ duration) fluctuations in~$D_0$, up to the level of about a factor of two, so that this $D(\gamma)$ movement is not entirely monotonic.
Overall, the values of $D_0$ are higher for higher-$\sigma_0$ simulations.
However, after about five \Alfven{} crossing times, the $D_0(t)$ lines for $\sigma_0 \geq 3/8$ simulations significantly overlap.
Meanwhile, those for $\sigma_0 \leq 3/32$ are more distinctly separated.

\begin{figure}
\centering
\includegraphics[width=\linewidth]{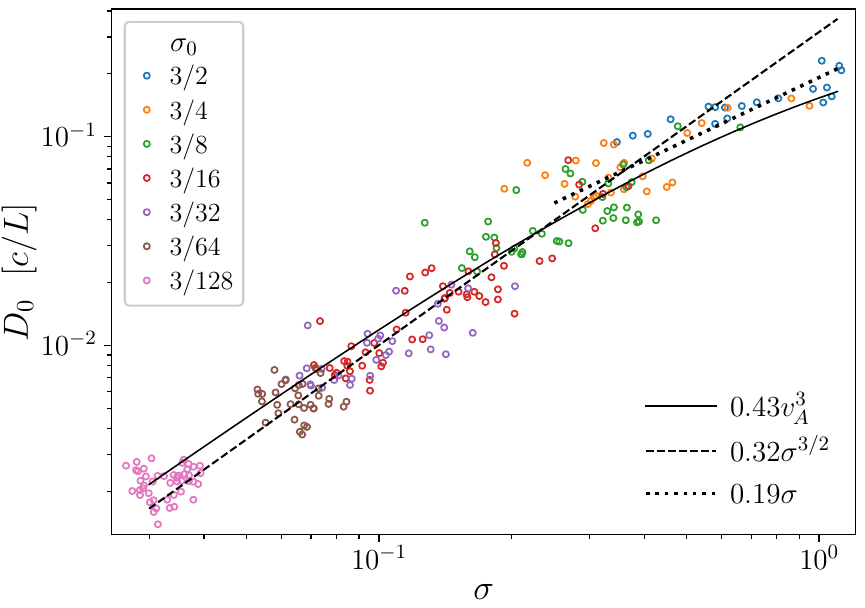}
\caption
{
The diffusion coefficient prefactor $D_0$ versus instantaneous magnetisation $\sigma(t)$ from all simulations in the \medSigScan{}, along with lines of $D_0 \propto \vA^3$ (solid), $D_0 \propto \sigma^{3/2}$ (dashed), and $D_0 \propto \sigma$ (dotted). The $D_0(\sigma)$ scaling is roughly $D_0 \propto \sigma^{3/2} \sim (v_A/c)^3$ at low $\sigma \lesssim 0.5$ and flattens somewhat at higher~$\sigma$, to around $D_0 \propto \sigma$.
}
\label{fig:d0vsSigma}
\end{figure}

%% D0(sigma)
Cross-referencing the values of $D_0(t)$ and $\sigma(t)$, we obtain~$D_0(\sigma)$, and plot this in \autoref{fig:d0vsSigma}.
We exclude points earlier than two \Alfven{} crossing times, before which turbulence has not fully developed. 
We also exclude points after the normalised system size $L/2 \pi \rho_e(t)$ drops below~15,
the size needed to start to see the inertial range and obtain converged particle distributions in our previous radiative turbulence study \cite{zhdankin2020KineticTurbulenceShining} (while that study states that $L/2 \pi \rho_e$ should be greater than~25, slightly lower values are acceptable for practical purposes). 

Here, we take advantage of the time dependence resulting from our simulation setup.
As the time evolution of $\sigma(t)$ from simulations with different $\sigma_0$ overlap, we ask whether $D_0$ also coincides at these times.
Indeed, we find that $D_0$ mainly depends on instantaneous $\sigma(t)$ rather than~$\sigma_0$. This is primarily supported by the points at $\sigma \gtrsim 0.1$ from simulations with $\sigma_0 \geq 3/16$, but some overlap is also seen for points with $\sigma_0 \geq 3/64$. For this range of simulations, the points with coinciding $\sigma(t)$ overlap even though they come from simulations with different~$\sigma_0$.
At the lowest $\sigma$, however, there is no overlap between different simulations. This is because the $\sigma(t)$ time evolution is much slower for lower~$\sigma_0$. Future work may benefit from more tightly spaced simulations in the low-$\sigma_0$ range.

We find the $D_0(\sigma)$ scaling to be consistent with $D_0 \propto \sigma^{3/2}$ at low $\sigma \lesssim 0.5$ (i.e., in the weakly relativistic regime), with the dependence flattening somewhat at high $\sigma \gtrsim 0.5$, to about $D_0 \propto \sigma$.
\autoref{fig:d0vsSigma} illustrates this behaviour with two fits: $D_0 = 0.43 (\vA/c)^3\, c/L = 0.43 [\sigma/(1+\sigma)]^{3/2}\, c/L$ and $D_0 = 0.32 \sigma^{3/2}\,c/L$; we see that they agree with the data very well at low~$\sigma \lesssim 0.5$, and provide a reasonably good agreement over the entire $\sigma$-range, although the $D_0 \propto \sigma^{3/2}$ scaling deviates noticeably at high~$\sigma$.
%(the quantity $\uA^2 \equiv \vA^2 / (1 - \vA^2/c^2) = \sigma$ is commonly used in Fermi acceleration theories).
Thus, our numerical results provide clear evidence against the standard Fermi-acceleration theory's prediction of a linear $D_0 \propto \sigma$ scaling for $\sigma \lesssim 0.5$, although this scaling may still be applicable for $\sigma \gtrsim 0.5$.
Our preferred scaling $D_0 \propto \vA^3$, which becomes $D_0 \propto \sigma^{3/2}$ in the low-$\sigma$ (nonrelativistic turbulence) limit, is consistent with the theoretical prediction by \citet{demidem2020ParticleAccelerationRelativistic} for \Alfven{ic} turbulence, in which the extra power of~$\vA$ (compared to the standard Fermi-acceleration theory) was attributed to resonance broadening.
Thus, our results provide direct numerical support for this theory. 
It is interesting to note, however, that the $D_0 \propto \vA^3$ scaling was predicted by \citet{demidem2020ParticleAccelerationRelativistic} only for Alfv\'enic turbulence, which is essentially incompressible, whereas for the case of turbulence dominated by compressional fast and slow magnetosonic waves that study obtained a traditional $D_0 \propto \vA^2$ scaling.  
We speculate that perhaps the reason why the $D_0\propto \sigma^{3/2}$ scaling is observed only at low~$\sigma$ in our simulations is that only in this nonrelativistic regime the turbulence is predominantly subsonic ($v_{\rm rms} \sim v_A \ll c_s \sim c$, where $v_{\rm rms}$ is the rms fluid bulk velocity and $c_s$ is the speed of sound) and hence essentially  incompressible, mediated mostly by the Alfv\'enic cascade. In contrast, at higher $\sigma$ turbulent motions are relativistic ($v_{\rm rms} \sim v_A \sim c_s \sim c$) and inevitably compressive, with a greater (perhaps even dominant) role played by fast and slow magnetosonic waves, which leads to the restoration of the traditional $D_0\propto \sigma$ scaling.
In principle, more simulations at higher $\sigma$ would help test this hypothesis and elucidate the high-$\sigma$ scaling; in practice, however, the high-$\sigma$ regime (with strong fluctuations, $\delta B_{\rm rms}/B_0 \sim 1$ and $v_{\rm rms} \sim c$) is difficult to maintain for any substantial period of time, because the magnetic energy would rapidly convert to thermal energy, quickly decreasing $\sigma$ before the turbulence can fully develop \citep{zhdankin2018NumericalInvestigationKinetic}.

Note that the normalization $D_0 (L/c)/(v_A/c)^3 \approx 0.43$ of the scaling fit in \autoref{fig:d0vsSigma} can be compared to the expectations from \citet{demidem2020ParticleAccelerationRelativistic} (their Eq. 38), which evaluates (up to a correction that is logarithmic in scale separation) to $\approx 0.6$ for a purely \Alfven{ic} cascade with amplitude $\delta B_{\rm rms}/B_0 \approx 1$. Thus, the results are in reasonable agreement. A precise theoretical comparison would require measuring the partition between different modes (\Alfven, fast, and slow) and taking into account corrections due to finite system size.

\subsection{System-size dependence}
\begin{figure}
\centering
\includegraphics[width=\linewidth]{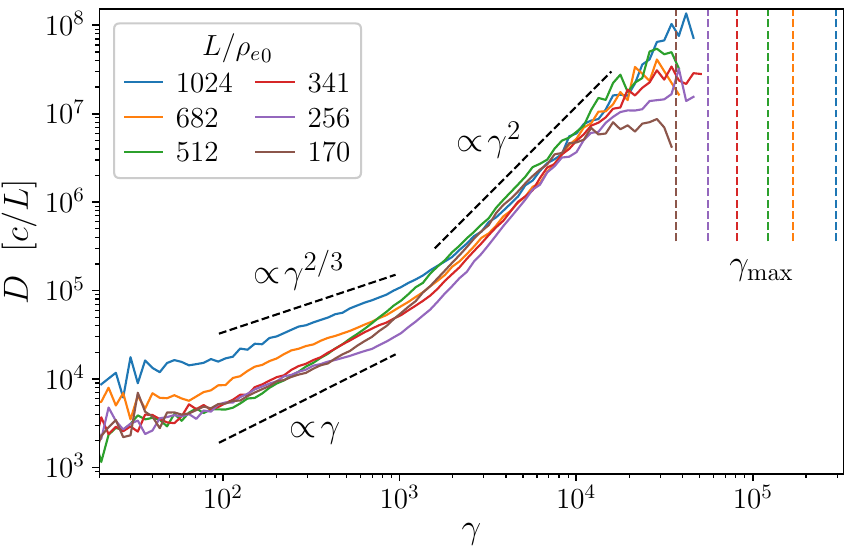}
\caption
{
Diffusion coefficient $D(\gamma)$ at $t_0 = 10L/c$ for each simulation in the system-size scan, where $\sigma_0 = 3/8$. Reference power-law scalings are shown in black dashed lines, and dashed vertical lines indicate $\gamma_{\rm max}$ for each case.
}
\label{fig:DvsL}
\end{figure}

%% D(x) L scan
\autoref{fig:DvsL} shows the diffusion coefficient at a single point of time $t_0 = 10L/c$ for the system-size scan simulations, which have the same $\sigma_0 = 3/8$ and varying $\relLar \in \{\relSizeVals\}$.
One may also choose time points with a logarithmic dependence on system size, in accordance with the convergence time in \citet{zhdankin2018SystemsizeConvergenceNonthermal}, but the difference is minor.
The agreement in the nonthermal range between the simulations of different $\relLar$ is good, showing uniform $D \propto \gamma^2$ scaling.
This indicates that $D(\gamma)$ in the general parameter space explored in this paper is reasonably well-converged with respect to $\relLar$ for the nonthermal region, providing confidence in our results. With regards to the extent of the $D \propto \gamma^2$ scaling, we would naively anticipate it to be proportional to the inertial range, and hence to the system size~$\relLar$. However, \autoref{fig:DvsL} shows that the energy range for which there are good $D$ measurements is almost unchanged with~$\relLar$. This is because the highest energy for which $D$ can be measured reliably in our study is limited by tracked particle statistics due to the $f \propto \gamma^{-\alpha}$ falloff rather than by the extent of the inertial range. 

The low-energy behaviour, roughly below $\gammaPeak \sim 10^3$, varies more substantially between the different~$\relLar$, but is  still broadly consistent with having a shallower power-law scaling than in the nonthermal range.
There is also a weak trend of higher-$\relLar$ values corresponding to larger low-energy diffusion coefficients.
The location of the spectral break between the two power laws also varies somewhat and does not show a particular pattern, but is roughly at $\gamma\sim 10^3$, a little bit above~$\gammaPeak$.
We leave further investigation of the low-energy range to future work. 

\begin{figure}
\centering
% \includegraphics[width=\halfwidth]{20230512_dvt_d0vsTime.pdf}
% \includegraphics[width=\halfwidth]{svt_sigmaVsT_s1d2.png}
% \includegraphics[width=\linewidth]{20230606_p2lscdsg_d0tsigtLscan.pdf}
% \begin{overpic}[width=\linewidth]{20240427_p2lscdsg_d0tsigtLscan.pdf}
\begin{overpic}[width=\linewidth]{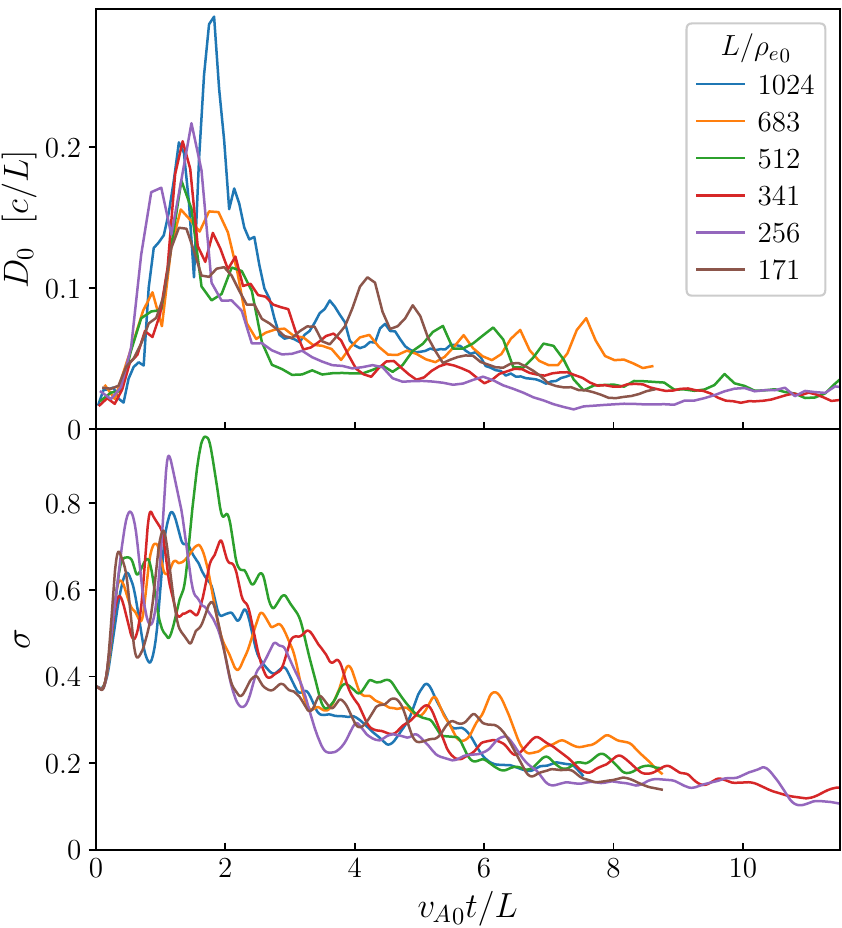}
    \put (50, 95) {(a)}
    \put (50, 49.5) {(b)}
\end{overpic}
\caption
{
Time evolution of (a) diffusion coefficient prefactor $D_0(t)$ and
(b) magnetisation $\sigma(t)$,
for the system-size scan,
where the simulations have the same $\sigma_0 = 3/8$ but different $\relLar \in \{\relSizeVals\}$. There is no clear systematic variation of these quantities with system size.
}
\label{fig:d0tsigtLscan}
% \label{fig:d0vstLscan}
% \label{fig:sigmaVsTLscan}
\end{figure}

%% sigma(t) and D0(t) L scan
\autoref{fig:d0tsigtLscan} presents $D_0(t)$ and $\sigma(t)$ for the system-size scan, showing that $\relLar$ has only a weak effect on the respective quantities, with lines from different simulations largely overlapping amidst the fluctuations. However, it is hard to tell if this is true due to time fluctuations in the traces.
Repeated runs of simulations with the same initial system parameters but different random seed would be required to test if there is any statistically significant difference in $D_0(t)$ or $\sigma(t)$ for simulations between which only the system size~$L$ varies.
We leave such an investigation for future work.
As in \autoref{subsec:Dsigscan}, the $D_0(t)$ and $\sigma(t)$ are qualitatively similar to each other, including the initial increase, subsequent decay, and dependencies on $\sigma_0$ and~$N$.

\subsection{Particles-per-cell convergence}
\begin{figure}
\centering
\includegraphics[width=\linewidth]{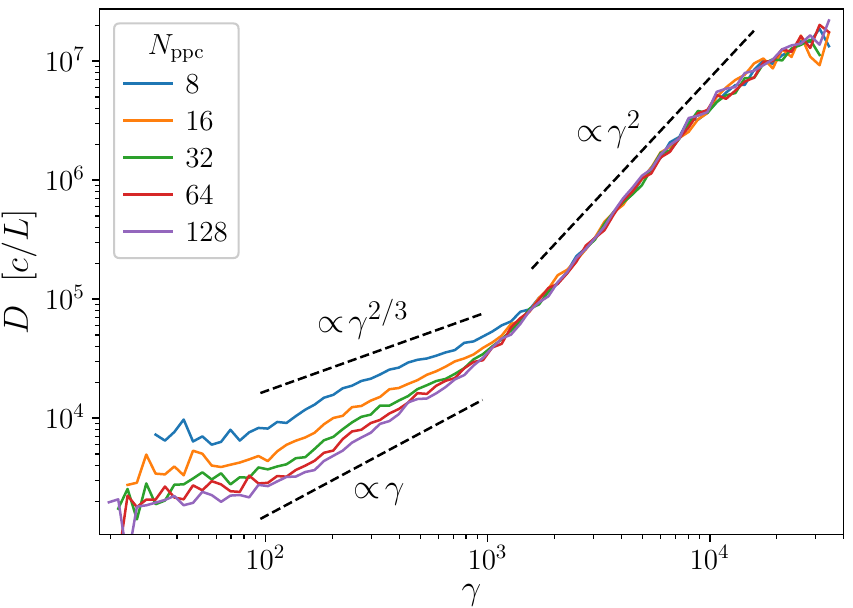}
\caption
{
Diffusion coefficient $D(\gamma)$ at $t_0 = 10L/c$ for the $\Nppc$-scan simulations, which have $\relLar = 256$ and $\sigma_0 = 3/8$.
In the nonthermal region, there is excellent convergence. At low energy, $D$ is converged for simulations of high $\Nppc$, but increased by numerical effects when  $\Nppc$ is too low.
}
\label{fig:DvsPPC}
\end{figure}

\autoref{fig:DvsPPC} shows the dependence of $D(\gamma)$ at fixed $t_0=10 L/c$ on the number of particles per cell $\Nppc$ for simulations with $\sigma_0 = 3/8$ and $\relLar=256$ ($N = 384$). The convergence in the high-energy, $D \propto \gamma^2$, range is excellent. The extent of this range shortens slightly as $\Nppc$ decreases, with the beginning occurring at slightly higher~$\gamma$. 
Lower energies also show a power-law scaling, which becomes shallower with decreasing~$\Nppc$. These two trends are displayed consistently and monotonically with respect to~$\Nppc$. Importantly, even this lower-energy segment shows convergence with increased~$\Nppc$, with essentially full convergence achieved for $\Nppc \gtrsim 64$ for the entire range of particle energies.
Note that the effect of decreasing $\Nppc$ in \autoref{fig:DvsPPC} resembles the trend of increasing system size from \autoref{fig:DvsL}; this resemblance may be explained by PIC noise having a stronger effect on small scales (low-energy diffusion) in simulations with a larger inertial range.

\subsection{Summary}
The diffusion coefficient as a function of particle energy $D(\gamma$) has broadly similar behaviour in the nonthermal range over simulations with different
$\sigma_0$, $L/\reo$, and~$\Nppc$.
The high-energy scaling of $\smash{D \propto \nobreak \gamma^2}$ is universal, while the low-energy scaling varies, with the strongest qualitative dependence on~$\sigma_0$.
The prefactor $D_0$ of the $\smash{D \propto \nobreak \gamma^2}$ scaling depends mostly on instantaneous magnetisation $\sigma(t)$ rather than~$\sigma_0$. 
For $\sigma \lesssim 0.5$, which is most of our explored range, the $D_0$ scaling is consistent with $D_0 \propto \vA^3$ or $D_0 \propto \sigma^{3/2}$, in contrast with the common theoretical prediction of $D_0 \propto \sigma$.
The convergence of $D$ in the nonthermal range with increasing system size $L/\reo$ and $\Nppc$ is excellent, which gives us confidence in the rest of our results.

%\clearpage
\section{Advection coefficient} \label{sec:aCoeff}
In \citet{wong2020FirstprinciplesDemonstrationDiffusiveadvective}, we observed that the particle energy-advection coefficient~$A$ was positive at $\gamma < \gammaPeak$ and negative at $\gamma > \gammaPeak$. However, the data at highest nonthermal energies in that study seemed too noisy to analyse in detail.
Now, our model in \autoref{sec:analyticalModel} suggests a concrete energy dependence $\anaAdvEqn$ \eqref{eqn:Axlogx} in the nonthermal range, and also predicts relationships between the \ac{FP} coefficient model parameters and the power-law index. The present Section examines the advection coefficient in the context of that model.

\begin{figure*}
\centering
\includegraphics[width=\linewidth]{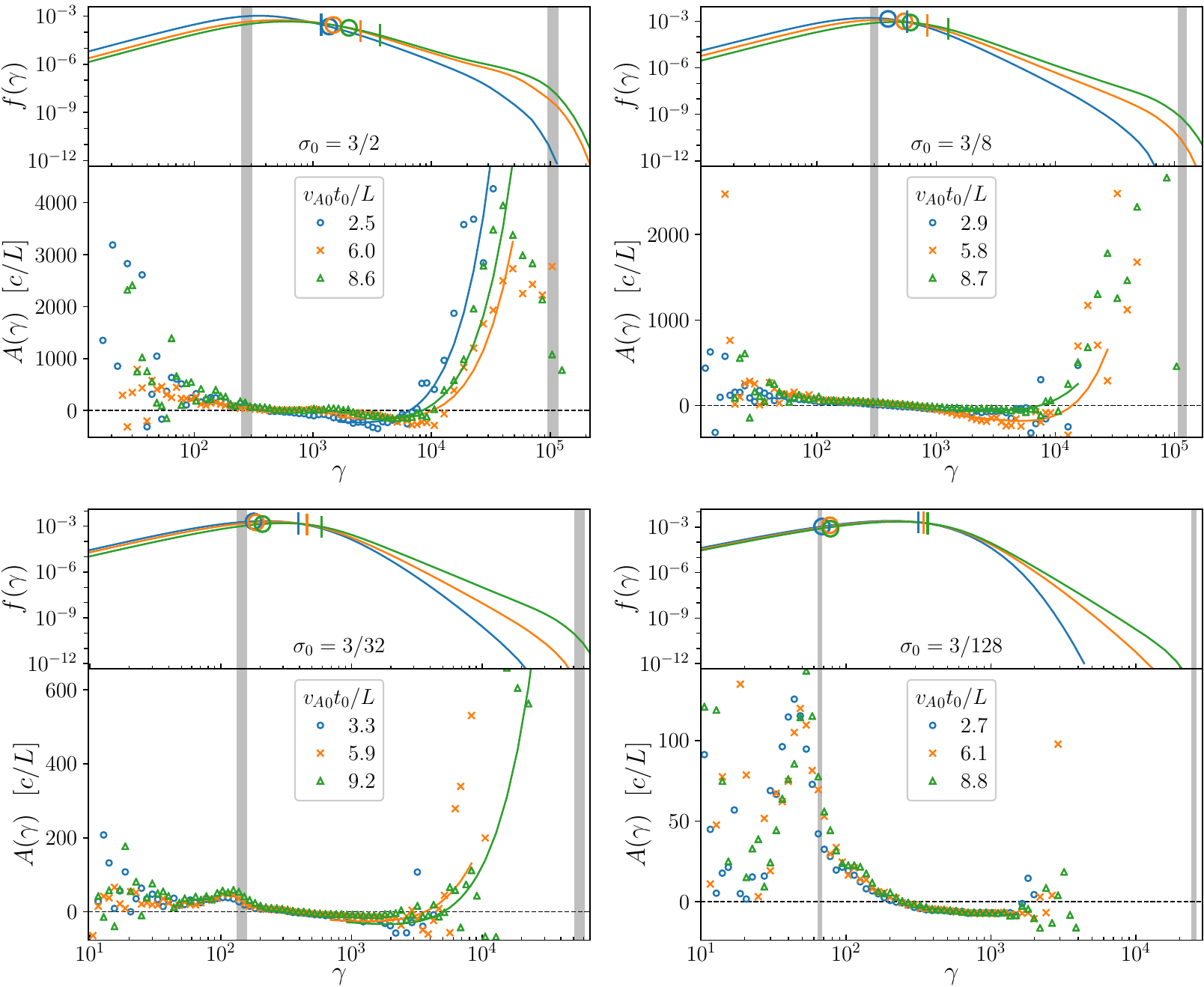}
\caption
{
Lower subpanels show the energy dependence of the advection coefficient $A(\gamma, t)$ (markers) at three different times (shown in the legend) for PIC simulations with $\sigma_0 \in \{3/2, 3/8, 3/32, 3/128\}$ (as labelled in each panel), with fits of $\anaAdvEqn$ (solid lines). Noise at high energies has been reduced by combining every two bins after $\gamma=10^4$ and averaging $A$ over a time $0.5 L/\vao$.
Upper subpanels show the particle energy distribution at corresponding times, for reference.
See the \autoref{fig:Dxt} caption for a description of the energy-scale markings.
}
\label{fig:Axt}
\end{figure*}

First, we reduce the noise in $A$ by time-averaging it over intervals of $0.5 L/\vao$ and also by merging consecutive pairs of bins above $\bcGamma=10^4$.
\autoref{fig:Axt} shows the resulting $A(\gamma, t)$ at times \figXtTimes{} for the simulations with $\sigma_0 \in \{\sigmaFourVals\}$ from the \medSigScan{}.
Although still considerably noisy in the nonthermal range, the qualitative behaviour is clear and is also similar throughout the depicted simulations and time instants:
$A$ is positive at highest and lowest energies, and negative in an intermediate interval.
The highest-energy behaviour is not visible if the spectral power law is too steep to have enough high-energy tracked particles to measure $A$ there; this occurs for the $\sigma_0 = 3/128$ simulation and at early times in the $\sigma_0 = 3/8$ and $\sigma_0 = 3/32$ simulations.
However, in these cases the lower-energy data still follows the aforementioned pattern.
The first zero crossing, from positive to negative near~$\gammaPeak$, was the initial observation of \citet{wong2020FirstprinciplesDemonstrationDiffusiveadvective}: that the advection coefficient tends to gather particles together in energy space.
This applies to thermal energies and so out of the \autoref{sec:analyticalModel} model's validity range.
The second zero crossing, from negative to positive, is a key feature of the \autoref{sec:analyticalModel} model that is borne out in these simulation results. This crossing occurs in the power-law range, but we note that the model does not predict nor require anything about its location.
This sign change is unlikely to be an artefact of the particle pileup at the system size limit $\gammaMax$ because it occurs at energies roughly an order of magnitude lower than $\gammaMax$ (which is depicted as the rightmost grey bar in \autoref{fig:Axt}).

\autoref{fig:Axt} also shows fits of $A(\gamma)$ to \eqref{eqn:Axlogx}, limited to the nonthermal range, for the cases with sufficient high-energy data as identified above.
The fits are consistent with the measurements inside the model's validity range of the power-law tail. Deviations occur outside that range, near $\gammaPeak$ and~$\gammaMax$, which is expected.

\begin{figure*}
\centering
\includegraphics[width=\textwidth]{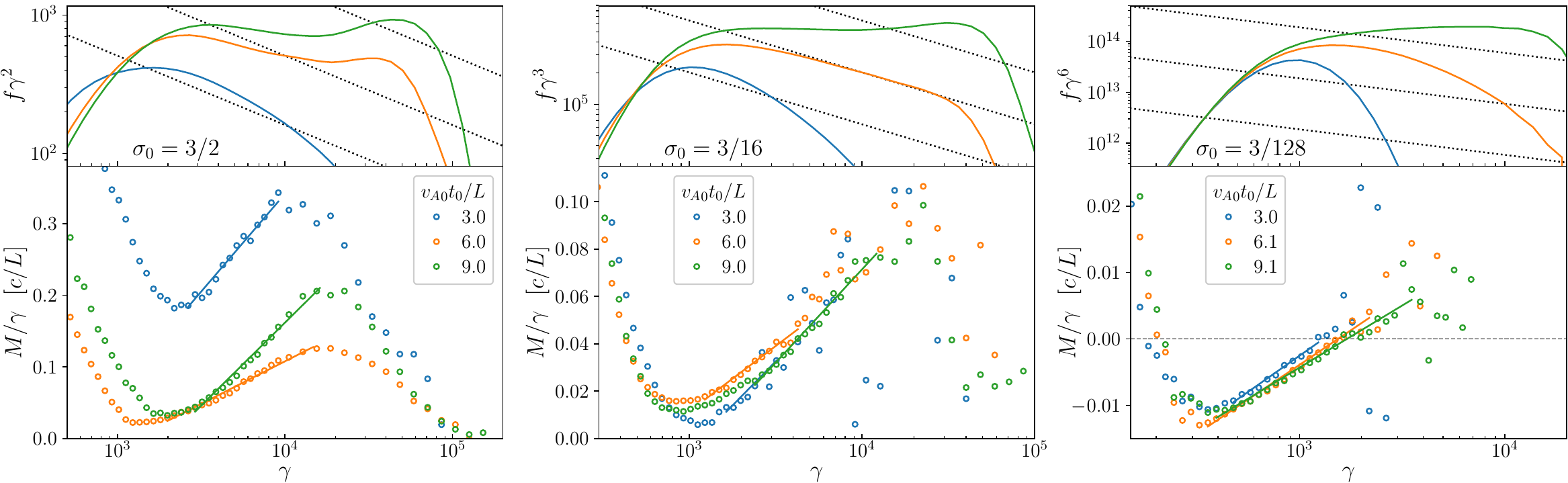}
\caption
{
Particle energy distribution (upper subpanels) and $M/\gamma$ (lower subpanels) as functions of energy $\gamma$ at three times $t \approx \{3,6,9\} L/v_{A0}$ for PIC simulations with different initial magnetisation $\sigma_0 \in \{3/2, 3/16, 3/128\}$ (from left to right).
Fits by $\log\gamma$ are shown in the lower subpanels (solid lines).
Upper subpanels are compensated by powers of energy such that the last particle energy distribution shown is roughly horizontal. Dotted lines are power laws with one-half lower index than the compensated value in each case, for reference.
}
\label{fig:dividedM}
\end{figure*}

We could now examine the trends and tendencies of the $A$ model equation \eqref{eqn:Axlogx} parameters~$A_0$ and~$\gStar_A$, and test their relationships (\ref{eqn:A0}--\ref{eqn:kIntA0}) with the power-law parameters $\alpha$ and~$K$.
However, we will first switch to the related variable~$M$, the energy-dependent average acceleration rate, defined in \eqref{eqn:meant}.
While the \ac{FP} equation \eqref{eqn:fpEne} is in terms of~$A$, it is more convenient to use $M$ because, by \eqref{eqn:meant}, $M$ comes from $\Delta\binMean$ measurements only, while $A =\nobreak M -\nobreak \pdx D$ is contaminated by noise in~$\pdx D$, with the $\gamma$-derivative particularly amplifying the noise.

Furthermore, we rewrite
$M = M_0 \gamma \log(\gamma/\gStar_M)$~(\autoref{eqn:Mxlogx})
into
    \begin{equation}
    M / \gamma = M_0 \log\gamma - M_0 \log\gStar_M.
    \label{eqn:divMlinx}
    \end{equation}
This now simply states that $M / \gamma$ should be linear in $\log\gamma$, with the $x$\nobreakdash-intercept at $\log\gStar_M$.
This form is convenient for visual inspection, and also
for fitting because linear regression is robust and well understood.

\autoref{fig:dividedM} shows $M/\gamma$ vs $\gamma$ on linear-log scales, so that a linear equation in $\log \gamma$ such as \eqref{eqn:divMlinx} appears as a straight line,
presented for a selection of simulations from the \medSigScan{} at measurement times \figXtTimes{}.
We also restrict the plot view to the high-energy region above~$\gammaPeak$.
In all cases, $M/\gamma$ has a remarkably good linear segment spanning about a decade of energies in the power-law region.
This is followed by a decline in $M / \gamma$ at energies near the system-size limit, adulterated by varying amounts of noise: less for high $\sigma_0$ and more for low~$\sigma_0$.
Surprisingly, even the lowest $\sigma_0 = 3/128$ simulation plot shows a high-quality straight-line region despite the paucity of high-energy particles (in contrast to the inconclusive nature of the corresponding region in \autoref{fig:Axt}).
The linear $M / \gamma$ region also appears clearly even before the $f(\gamma)$ power law has fully formed, as seen at the early $t_0 \approx 3L/\vao$ time instances.
The presence of this high-quality linear segment over a wide range of initial magnetisation values gives strong qualitative support to the \autoref{sec:analyticalModel} model.

\begin{figure}
\centering
\begin{overpic}[width=\linewidth]{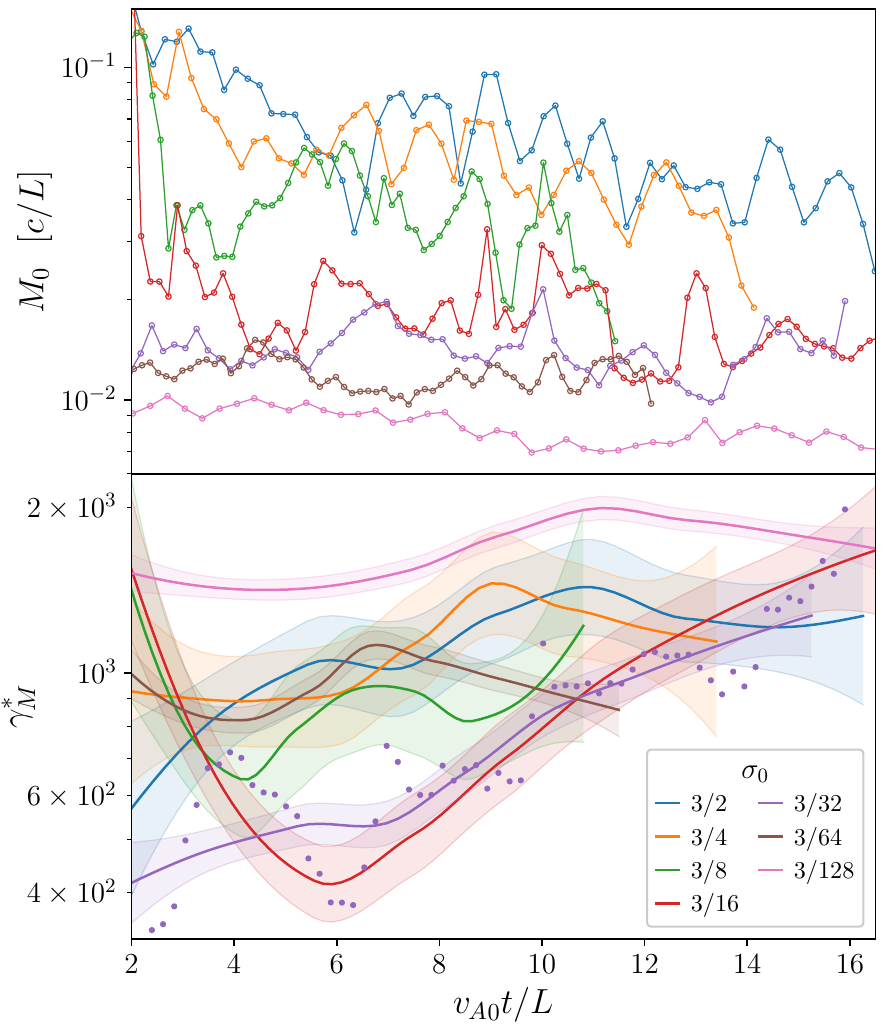}
    \put (30, 95) {(a)}
    \put (30, 49.5) {(b)}
\end{overpic}
\caption
{
(a)~%
Connected circles display $M_0$ obtained by fitting $M(\gamma)$ to~\eqref{eqn:divMlinx}, for varying $\sigma_0$ indicated in the legend. These show that $M_0$ is higher for higher $\sigma_0$ simulations, and also that it tends to decrease over time.
(b)~%
Solid lines are smoothed $\smash{\gStar_M}$ values (obtained by fitting $M(\gamma)$ to \autoref{eqn:divMlinx})
while the adjacent shaded regions indicate approximately the spread of unsmoothed measurements.
We display the original unsmoothed points for one simulation, to help visualise how the shaded areas relate to the point scatter.
We employ LOESS \citep{Cleveland1981LOWESSPrograSmooth} to generate the smoothed curves and confidence intervals, but this is a visual aid only, without any statistical purpose.
The $\smash{\gStar_M}$ values are around $\smash{\gamma = 10^3}$, which is typically in the early part of the nonthermal section, and $\gStar_M$ tends to increase over time, as one might expect from overall simulation heating.
}
\label{fig:0gStarVst}
\label{fig:compareCx}
\end{figure}

We now examine the coefficients $M_0$ and~$\gStar_M$, which we extract via \eqref{eqn:divMlinx} from the linear fits (as shown in \autoref{fig:dividedM}) to the straight-line region in~$M/\gamma$.
\autoref{fig:0gStarVst} displays $M_0(t)$ and $\gStar_M(t)$ for each simulation in the \medSigScan{}.
We find that $M_0$ is higher for higher-$\sigma_0$ simulations, and tends to decrease over time.
As $\gStar_M$ is too noisy to show all the points, we use LOESS \citep{Cleveland1981LOWESSPrograSmooth} to generate a smoothed curve, with shaded intervals to approximately indicate the point scatter. This is a visual aid only with no intended statistical meaning.
The $\smash{\gStar_M}$ values are around $\smash{\gamma = 10^3}$, which is typically in the early part of the nonthermal section, and $\gStar_M$ tends to increase over time, as one might expect from overall simulation heating.

\begin{figure}
\centering
\includegraphics[width=\linewidth]{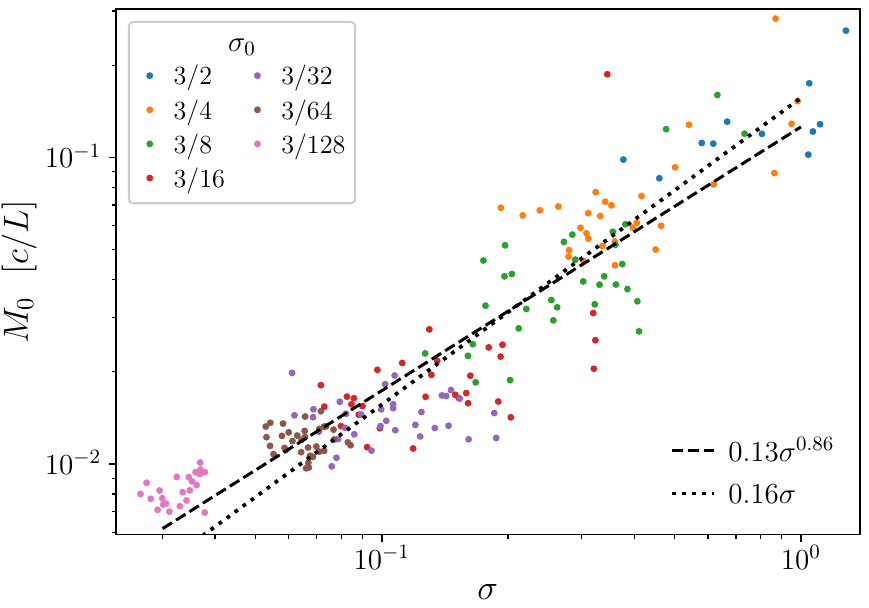}
\caption
{
The scaling of $M_0$ versus instantaneous magnetisation~$\sigma$, for PIC simulations with varying initial magnetisation $\sigma_0$ (indicated in legend), showing a positive correlation independent of $\sigma_0$. The relationship between~$M_0$ and~$\sigma$ is consistent with a linear scaling (particularly for $\sigma_0 \geq 3/64$; dotted line) or a slightly shallower power law ($\sigma^{0.86}$, dashed line).
}
\label{fig:M0sig}
\end{figure}

\autoref{sec:analyticalModel} does not obtain any specific predictions for $M_0$ and~$\gStar_M$ 
% or $A_0$ and $\smash{\gStar_A}$)
as functions of system parameters in the way that analytical models predict the scaling of (the diffusion-coefficient $\gamma^2$-prefactor) $D_0$ with instantaneous~$\sigma$ (see \autoref{subsec:Dsigscan}).
Nevertheless, by analogy to the diffusion coefficient, we plot $M_0$ against $\sigma$ in \autoref{fig:M0sig}.
The scaling is consistent with $M_0 \propto \sigma$, or slightly shallower.
The increasing trend matches the basic qualitative expectation from $M_0 =\nobreak A_0 =\nobreak - d\log{(\alpha-1)}/dt$ (\autoref{eqn:A0}): as we expect $\alpha$ to evolve faster at higher $\sigma$ (in order to reach harder power laws), $M_0$ should increase with~$\sigma$.
We leave more detailed interpretation to future investigation.

\subsection{Comparison to analytical model of power law evolution}
\autoref{sec:analyticalModel} derived relationships between the \ac{FP} coefficient parameters~$D_0$,
% $A_0$, and~$\gStar_A$
$M_0$, and~$\gStar_M$
and the power-law parameters $K$ and~$\alphaSV$.
These relations are the differential equations
% (\ref{eqn:A0}, \ref{eqn:xA}) and their equivalent integral forms (\ref{eqn:cInt}, \ref{eqn:kIntA0}).
(\ref{eqn:A0}, \ref{eqn:xM}) and their equivalent integral forms (\ref{eqn:alphaIntM0}, \ref{eqn:kIntM0}).
In this Subsection, we test the consistency of these formulae with the acquired data.
While equivalent, we prefer and use the integral forms.
This is because $A_0$ and $\gStar$ can both fluctuate significantly while still producing relatively smooth $\alpha(t)$ and~$K(t)$, because the latter two quantities are obtained through the cumulative effect of the \ac{FP} coefficients acting on the particle energy distribution.
That is, differential equation comparisons may mislead in terms of how much \ac{FP} coefficients' time fluctuations affect the power-law index.

The method for obtaining $D_0$ was described in \autoref{sec:dCoeff}; for $M_0$ and~$\gStar_M$, above; and for $\alphaSV$, in \autoref{sec:alpha}.
The procedure for finding the power-law prefactor~$K$ derives from extending the $\alphaSV$ measurement process.
That process identifies, at each instance of time, a preferred point on the local power-law index curve from which we obtain~$\alphaSV$.
To obtain $K$, we substitute that point's $(\gamma, f)$ coordinates into $f =\nobreak K \gamma^{-\alpha}$~\eqref{eqn:fPowerLaw} to yield $K =\nobreak f \gamma^\alpha$.
We note that $K$ suffers amplified noise from $\alphaSV$ fluctuations due to the long lever arm in its definition as $f(\gamma = 1)$, and future work may benefit from defining it relative to a more central point.

\begin{figure}
\centering
\includegraphics[width=\linewidth]{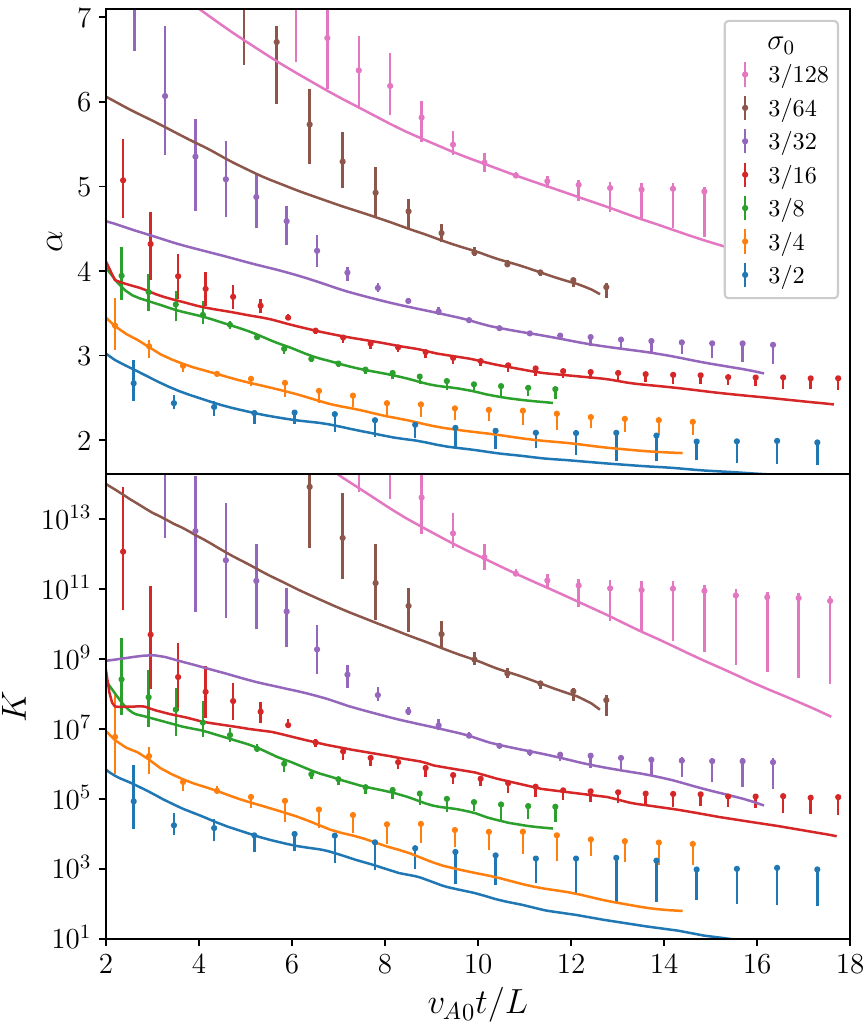}
\caption
{
Integral formulas computing $\alpha(t)$ and $K(t)$ from the fitted \ac{FP} coefficient model parameters $D_0$, $M_0$, and~$\gStar_M$ (solid curves) match reasonably well to $\alpha$ and $K$ measured directly from the particle energy distribution (points with vertical bars) at varying times.
The vertical bars are approximate possible ranges for each parameter due to the particle energy distribution local power-law slope having $\gamma$ dependence; see text for details.
}
\label{fig:xlxInts}
\end{figure}

\autoref{fig:xlxInts} compares $\alpha$ and $K$ measured from the particle energy distribution with those obtained through the time integrals (\ref{eqn:alphaIntM0}, \ref{eqn:kIntM0}) which are functions of the \ac{FP} coefficient model parameters $D_0$, $A_0$, and~$\gStar_A$.
This Figure displays vertical ``error bars" for $\alpha$ indicating a range of nearby values due to the energy dependence of the local power-law index $\alphaLocal(\gamma, t) \equiv \alphaDef$.
These $\alpha$ bars are the same as those in \autoref{fig:alphaExpSig} and their calculation method is specified in \autoref{sec:alphaLscan}.
The $K$ bars follow the same simple inversion procedure described just above, except using a range of $\alpha$ corresponding to the $\alpha$ bars.
We choose the integration constants so that the integral lines and $\alpha$ measurement lines intersect where the $\alpha$ ``error bar" is smallest. This corresponds roughly to where the power law is flattest, just before the high-energy pileup appears (see \autoref{sec:alpha}), which is essentially the ``inflection time" discussed by \citet{zhdankin2018SystemsizeConvergenceNonthermal}.
Substantially before this time, the power law is not fully formed, and so we focus mainly on the vicinity of this moment and later.

We find in general that the integral-derived curves are consistently steeper than the $\alphaSV$ and $K$ points measured from the power-law slope.
This might be explainable by examining the ``error bars". After the inflection point, these are consistently unidirectional because the \autoref{sec:fittingPlaw} $\alphaSV$-measurement method chooses a local extreme of $\alphaLocal$ and so nearby values are all to one side.
The bar direction is towards lower $\alpha$ magnitude, corresponding to a flatter power law; this is because including the high-energy pileup in any kind of averaging fit would decrease the apparent overall power-law index.
Hence, we observe that beyond the constructed intersection point, the \autoref{fig:xlxInts} integral curves all pass on the bar side of the $\alpha$ and $K$ points.
This then implies that the model of \autoref{sec:analyticalModel} is sensitive in part to the high-energy pileup, and has an effective power-law index somewhat smaller in magnitude than that obtained from the \autoref{sec:fittingPlaw} fitting. This could be further examined in future work by using a power-law fitting method that trims the high-energy pileup less aggressively.

These comparisons between the $\alpha$ and $K$ (power-law based) and \ac{FP} measurements stretch the data quality, with substantial uncertainty on both sides. The uncertainty for $\alpha$ is due to the limited extent of the power law, and for the \ac{FP} coefficients, due to the limited number of high-energy tracked particles.
Hence, these consistent results only indicate that the model is quantitatively plausible.
It is nevertheless significant because the \ac{FP} coefficients come from tracked particles while the power-law parameters come from the global particle energy distribution, two substantially different data acquisition methods.

Overall, this Section's results provide confidence in the basic validity of the \autoref{sec:analyticalModel} model.
In addition to the reasonable quantitative agreement just discussed, the remarkably clear linear segment in $M/\gamma$ vs. $\log\gamma$ in \autoref{fig:dividedM} gives strong qualitative support.
It is also encouraging that the assumptions underlying the \autoref{sec:analyticalModel} model are very simple.

%\clearpage
\glsresetall
\section{Conclusions}
\label{sec:conc}
In this paper, we investigated the evolution of nonthermal particle populations in first-principles \ac{PIC} simulations of driven relativistic pair plasma turbulence through the lens of the \ac{FP} energy diffusion-advection model.
Measurements of the power-law index $\alpha$ of the nonthermal particle energy distribution demonstrate behaviour consistent with an exponentially converging time evolution.
We then examine the behaviour of particle energies by tracking large numbers of simulated particles.
We demonstrate that particle energies behave diffusively with respect to time over broad ranges of particle energies and system parameters. 
We measure the energy diffusion and advection coefficients ($D$ and $A$) as functions of relativistic particle energy and time using the tracked-particle data.
When fed back into the \ac{FP} equation, these coefficients successfully reproduce the evolution of the particle energy distribution obtained from the \ac{PIC} simulations.
This confirms that the \ac{FP} model provides an adequate description of particle acceleration for the entire investigated system parameter space.
We find that the energy diffusion coefficient $D(\gamma)$ generally scales with energy $\gamma mc^2$(or momentum~$\gamma mc$) squared in the nonthermal range: $D(\gamma)= D_0 \gamma^2$. 
The coefficient of this scaling, $D_0$, depends primarily on the instantaneous magnetisation $\sigma$ rather than the initial magnetisation, and this relationship is consistent with $D_0 \propto (\vA/c)^3 \sim \sigma^{3/2}$ at low $\sigma \lesssim 1$. 
The interpretation of the advection coefficient measurements is guided by an analytical model relating the nonthermal power-law index to the \ac{FP} coefficients in the nonthermal range.
This predicts a scaling of $\anaAdvEqn$, which is borne out by the measurements.
In addition, quantitative relationships between parameters fitted to the \ac{FP}-coefficient variables and the index and normalisation of the power-law tail are fulfilled reasonably well.

This work opens various avenues for further exploration.
The diffusion coefficient scaling is somewhat different from the expected scaling of $D_0 \propto \sigma$. This could be further investigated through simulation and analytical theory.
The new scaling of $\anaAdvEqn$ has significant implications for the evolution of the nonthermal energy spectrum (as was shown in \autoref{fig:fpEvolution}).
This may be of interest as nonthermal particle acceleration models commonly assume that the momentum-space advection coefficient is zero (leading to $A = 2D/\gamma$).
While we have focused on the behaviour of the \ac{FP} coefficients in the nonthermal region, we also measured $D$ and~$A$ at thermal and subthermal energies, finding qualitatively different behaviour.
This low-energy range may influence nonthermal particle acceleration through the injection of intermediate-energy particles for stochastic acceleration, and may be highly sensitive to the intermittency of turbulence \citep[e.g.,][]{comisso_sironi_2019, vega_etal_2023, davis_etal_2024}. 
In any case, further investigation may shed light on other physical phenomena relevant at those energies and scales.

This study of magnetised turbulent nonthermal particle acceleration enhances our understanding of widespread and long-studied plasma physical processes, and has profound implications for space, solar, and astrophysical systems. Apart from turbulence, other plausible mechanisms of nonthermal particle acceleration include magnetic reconnection and shocks. Our analysis methods may be adapted to kinetic simulations of these different processes in order to test theories of particle acceleration in those respective environments.
In future work, our methods may also be applied to model the asymptotic particle distribution function in PIC simulations of turbulence with strong radiative cooling \cite{zhdankin2020KineticTurbulenceShining, zhdankin_etal_2021, groselj_etal_2024} or escaping particles, in which a true statistical steady state is achieved.

\section*{Acknowledgements}
The Authors wish to thank Martin Lemoine for encouraging and insightful discussions of turbulent particle acceleration. 
This work was primarily supported by NSF grant AST 1806084, which is gratefully acknowledged.  DU, MB, and GW also acknowledge partial support from NASA Astrophysics Theory Program grant 80NSSC22K0828. 
VZ acknowledges support from NSF grant PHY-2409316. 
This research was also supported in part by grant NSF PHY-2309135 to the Kavli Institute for Theoretical Physics (KITP), through the participation of VZ and DU at the KITP program on ``Interconnections between the Physics of Plasmas and Self-gravitating Systems.'' This work used Stampede2 at the Texas Advanced Computer Center (TACC) through allocation TG-PHY160032 from the Advanced Cyberinfrastructure Coordination Ecosystem: Services \& Support (ACCESS) program, which is supported by U.S. National Science Foundation grants \#2138259, \#2138286, \#2138307, \#2137603, and \#2138296.

\section*{Data Availability Statement}
The data underlying this article will be shared upon a reasonable request to the corresponding author.

\bibliographystyle{mnras}
% \bibliography{dbx_refs}
\bibliography{autoexport,refs_manual}

%% The below is from the mnras template and it says not to change it.
% Don't change these lines
\bsp	% typesetting comment
\label{lastpage}
\end{document}